\documentclass[12pt,preprint]{aastex}

\shorttitle{SCR Photometric Distance Estimates}
\shortauthors{Winters et al.}

\begin{document}

\title{The Solar Neighborhood XXIII CCD Photometric Distance Estimates
 of SCR Targets -- 77 M Dwarf Systems within 25 Parsecs}

\author{Jennifer G. Winters\altaffilmark{1}, Todd
J. Henry\altaffilmark{1}, and Wei-Chun Jao\altaffilmark{1}}

\affil{Department of Physics and Astronomy, Georgia State University,
Atlanta, GA 30302-4106}\email{winters@chara.gsu.edu,
thenry@chara.gsu.edu, jao@chara.gsu.edu}

\author{John P. Subasavage\altaffilmark{1}}

\affil{Cerro Tololo Inter-American Observatory, Casilla 603, La
Serena, Chile}\email{jsubasavage@ctio.noao.edu}

\author{Charlie T. Finch\altaffilmark{1}}

\affil{United States Naval Observatory, Washington DC
20392-5420}\email{finch@usno.navy.mil}

\and

\author{Nigel C. Hambly}

\affil{Scottish Universities Physics Alliance (SUPA), Institute of
Astronomy, University of Edinburgh, Royal Observatory, Blackford Hill,
Edinburgh EH9 3HJ, Scotland, UK}\email{nch@roe.ac.uk}

\altaffiltext{1}{Visiting Astronomer, Cerro Tololo Inter-American
Observatory.  CTIO is operated by AURA, Inc.\ under contract to the
National Science Foundation.}

\begin{abstract}

We present CCD photometric distance estimates of 100 SCR
(SuperCOSMOS RECONS) systems with $\mu$ $\geq$ 0$\farcs$18/yr, 28 of
which are new discoveries previously unpublished in this series of
papers. These distances are estimated using a combination of new $VRI$
photometry acquired at CTIO and $JHK$ magnitudes extracted from 2MASS.
The estimates are improvements over those determined using
photographic plate $BRI$ magnitudes from SuperCOSMOS plus $JHK$, as
presented in the original discovery papers. In total, 77 of the 100
systems investigated are predicted to be within 25 pc. If all 77
systems are confirmed to have $\pi$$_{trig}$ $\ge$ 40 milliarcseconds,
this sample would represent a 23\% increase in M dwarf systems nearer
than 25 pc in the southern sky.

\end{abstract}

\keywords{stars: distances --- stars: low mass --- stars: statistics
  --- solar neighborhood --- techniques: photometric}

\section{Introduction}
\label{sec:intro}

For more than a century, there has been serious effort invested in
compiling a complete survey of stars near the Sun. Recent work of note
includes L{\'e}pine's SUPERBLINK program that has detected dwarfs,
giants, and subgiants in the northern sky with spectroscopic distances
$\leq$ 25.0 pc (L{\'e}pine 2005b) and additional objects in the south,
but without spectral types or distance estimates (L{\'e}pine 2005a;
L{\'e}pine 2008).  Reid and collaborators have also made progress in
this area, focusing on objects within 20 pc using spectrophotometry,
as reported in the Meeting the Cool Neighbors papers (Reid et al.~2008
and references therein). These efforts include both previously known
and new members of the solar neighborhood.  The survey most similar to
our survey discussed here is that of Deacon who used SuperCOSMOS $I$
plates and 2MASS to reveal new nearby red objects in the southern sky
(Deacon et al.~2005, 2007). This paper focuses specifically on new
discoveries from our SuperCOSMOS RECONS (SCR) effort (Hambly et
al.~2004, Henry et al.~2004, Subasavage et al.~2005a, Subasavage et
al.~2005b, Finch et al.~2007).

The most widespread method of revealing new members in the Sun's
vicinity is through proper motion searches, as nearby objects
generally have larger proper motions than those further away.  Much of
the nearby star work in the latter half of the twentieth century is
based upon the Palomar and UK Schmidt sky surveys done by Luyten
(Luyten 1979a, 1979b) and the Lowell proper motion surveys by Giclas
(Giclas et al.~1971; Giclas et al.~1978a, 1978b). Long-term efforts to obtain
optical photometry by Weis (Weis 1996) in the northern hemisphere and
Eggen (Eggen 1987) in the southern hemisphere were carried out by
observing one star at a time.  This photometry is often used to
estimate distances to stars using various color-absolute magnitude
calibrations.  Much of these data were then incorporated into the
Catalog of Nearby Stars (CNS, Gliese and Jahrei\ss ~1991), which
provided a snapshot view of the solar neighborhood population, albeit
one that was incomplete.  NStars continued in this vein with a
comprehensive database of objects with trigonometric parallaxes (see
Henry et al.~2002 and Henry et al.~2003 for more complete
discussions.). As we entered the twenty-first century, large sky
surveys such as 2MASS (Skrutskie et al.~2006), DENIS (The Denis
Consortium 2005), and SDSS (Adelman-McCarthy et al.~2009) gathered
magnitudes in optical and infrared filters of huge numbers of sources
that could be examined for nearby star candidates.

Once a new nearby star candidate is found, distances can be determined
using a variety of astrometric, photometric, and spectrophotometric
techniques.  In this paper, we focus on distance estimates made using
photometry from a combination of optical CCDs and infrared arrays,
which is an effective intermediate method for refining membership of
the solar neighborhood. This method provides distances that are more
accurate than those determined using photographic plates, but that are
not as precise (or as time and labor intensive) as those calculated
via trigonometric parallax.

A primary goal of the RECONS group (the Research Consortium on Nearby
Stars\footnote{\it www.recons.org}) is to identify hidden members of
the solar neighborhood within 25 pc to match the horizons of the CNS
and NStars efforts.  We have focused our searches in the southern
hemisphere because that portion of the sky has been historically
underrepresented and is expected to yield rich treasures in the form
of new stellar neighbors.  In this paper, we first discuss recent
photometric distance efforts by others (\S 2), then provide
descriptions of our two methods used to estimate distances
photometrically: via scanned photographic plates (\S 3) and
improvements based on CCD photometry (\S 4), with complementary
spectroscopy (\S 5).  We sum up the results in \S 6.

\section{Recent Photometric Distance Efforts by Others}
\label{sec:photparallaxes.others}

Table 1 lists the main photometric distance estimation efforts in the
southern sky since the NStars Database was compiled. The efforts
considered are purely photometric and are independent of any spectral
types or relations based on spectroscopy. The criteria for inclusion
in the counts are that the systems (1) be new discoveries, (2) have
colors of red dwarfs (effectively spectral type M, ($V-K$) $\gtrsim$
3.0), (3) be found in the southern hemisphere (DEC $\leq$~0$^\circ$),
and (4) are estimated to be within 25.0 pc via a photometric distance
estimate.

We used the NStars Database as a benchmark to evaluate solar
neighborhood completeness, as NStars incorporated only those objects
for which a parallax had been measured; therefore, it was the most
definitive collection of systems found within 25 pc. Great care was
taken when extracting the sample of 329 southern M dwarf systems from
NStars. All northern hemisphere objects, as well as those objects
brighter than $M_V$ $=$ 9.00, the adopted cutoff for the brightest M
dwarfs, were eliminated.  For the remaining objects with no designated
spectral type, various colors such as ($V-I$), ($V-J$), and ($J-K$),
depending on the availability of photometry, were used to eliminate
white dwarfs that remained in the sample. If the primary
component of a multiple system was discovered to be anything other
than an M dwarf, or if the system contained a white dwarf, the system
was omitted. Finally, the individual components of multiple systems
were combined to give a count of southern M dwarf systems.

Figure 1 shows results from these photometric distance efforts in
concert with the 329 systems from NStars (small dots). Declination is
plotted against Right Ascension in a polar plot with RA progressing
counterclockwise from 0 to 24 hours and DEC starting at 0$^\circ$ at
the outer boundary and moving inward to $-$90$^\circ$ at the
center. Phan-Bao's 37 objects\footnote{In the case of Phan-Bao et
al.~2001, where two different distances were given, we used the
Dist$_I$ that they considered more dependable. Their photometric
distances were estimated using DENIS $I-J$ colors with errors
$\sim$15-45\%.} are denoted by open triangles, Reyl\'e's 34
objects\footnote{In the case of Reyl\'e et al., we used only those
distance estimates that were unambiguously within 25.0 pc, i.e., we do
not count those stars with multiple estimates in Reyl\'e et
al.~2002. Their photometric distances were estimated using DENIS $I-J$
colors with errors up to 45\%.} are denoted by open circles, Finch's
15 objects\footnote{Finch's photometric distances are based on
SuperCOSMOS $BRI$ and 2MASS $JHK$ photometry and were estimated using
a method identical to that discussed in this paper for plate
photometry estimates.} are noted by upside-down open triangles, and
Costa \& M\'endez's three objects\footnote{The photometric distances
were estimated using $M_V$ vs.~$V-R$ and $M_V$ vs. $V-I$ relations
derived in Henry et al. 2004, with errors $\sim$15\%.} are indicated
with open squares. The 77 systems found to date in SCR surveys
(discussed in the next section) are represented by solid stars.  In
all cases, the counts and points in Figure 1 represent newly
discovered southern systems.

\section{Definition of the SCR Sample}
\label{sec:scrsampledef}

SCR objects are previously undiscovered systems that have been
revealed through searches utilizing the digitally scanned SuperCOSMOS
photographic plates. A sample of 72 star systems was harvested from
five earlier papers in this series (Hambly et al.~2004, Henry et
al.~2004, Subasavage et al.~2005a, Subasavage et al.~2005b, Finch et
al.~2007). To date, an additional 28 systems have been found during
continuing searches and are reported here for the first time. Table 2
lists the names, coordinates, and SuperCOSMOS $B_J$, $R_{59F}$, and
$I_{IVN}$ plate magnitudes. Additional data for these objects are
found in Table 3.

The 100 SCR systems (104 objects) listed in Table 3 comprise all SCR
red dwarfs with proper motions $\geq 0\farcs18$/yr estimated to be
within 25.0 pc based upon ``plate distance estimates.''  These
estimates use SuperCOSMOS $B_J$, $R_{59F}$, and $I_{IVN}$ plate
magnitudes (hereafter $BRI$) and $JHK_{s}$\footnote{Henceforth, the
subscript on the 2MASS $K_{s}$ filter will be dropped.} magnitudes
from 2MASS.  The 100 systems all have M dwarf primaries, two of which
are binaries with separations large enough to permit separate plate
distance estimates for the secondaries, SCR 2241-6119B and SCR
2335-6433B.  Two additional objects, SCR1107-3420B and SCR1746-8211,
have CCD distance estimates that place them within 25.0 pc, but they
are not included in the counts of M dwarf discoveries because the
primary of the system is not an M dwarf.

In Table 3 columns 1--3 are the names, RA, and Dec (epoch and equinox
2000.0). Columns 4 and 5 are the proper motions and position angles as
measured by SuperCOSMOS.  Columns 6--9 list the $VRI$ magnitudes and
the number of nights each system was observed.  The 2MASS $JHK$
magnitudes are listed in columns 10--12. Columns 13--18 provide the
plate distance estimates, related errors, and number of color-absolute
magnitude relations utilized (see next paragraph), followed by the CCD
distance estimates (see $\S$4.3), the corresponding errors, and number
of relations utilized. The tangential velocities, based on the CCD
distance estimates and the SuperCOSMOS proper motions, are given in
column 19.  Column 20 lists the spectral types, where available, and
column 21 includes the references for the publications that originally
reported each object, where the 29 new objects in this paper are noted
with reference 6. A final column (22) is provided for notes.

A system's plate distance estimate determines its membership in this
sample and is based on up to 11 useful M$_K$ vs.~color relations, as
described in Hambly et al.~(2004).  Relevant errors in the SuperCOSMOS
plate photometry for this sample, based on the ranges in $BRI$
magnitudes for the sources, are 0.2$-$0.6 mag for $B$ $=$ 11$-$21,
0.1$-$0.3 mag for $R$ $=$ 9$-$18, and 0.2$-$0.4 mag for $I$ $=$ 8$-$16
(Hambly et al.~2001). The errors on the individual passbands are
largely systematic in origin and are correlated between the passbands
because of the calibration technique used; however, the errors on
colors like ($B-R$) and ($R-I$) are smaller, which is more important
for their use in distance-color relations (see also $\S$4.4).

A plate distance estimate is considered reliable if all 11 relations
are applicable, i.e.~if~a star's color falls within the range covered
by the calibrations detailed in Hambly et al.~(2004) for single, main
sequence stars.  Between six and ten relations indicates that the
distance estimate is indicative, but less reliable.  Six relations are
adopted because if one plate magnitude is erroneous, up to five
relations may drop out.  Thus, if at least six relations are valid,
then at least two of the three $BRI$ magnitudes provide
optical/infrared colors consistent with normal main sequence stars.
In the present sample, only two stars have fewer than six valid
relations, so their distance estimates are highly suspect.  In fact,
the improved distance estimates discussed next indicate that both are
beyond 50 pc.

\section{Improving the Distance Estimates}
\label{sec:phot}

\subsection{$VRI$ CCD Photometry}

Our goal was to obtain at least two sets of $V_{J}$, $R_{KC}$ and
$I_{KC}$\footnote{Subscript: J $=$ Johnson, KC $=$ Kron-Cousins. The
central wavelengths for $V_{J}$, $R_{KC}$ and $I_{KC}$ are 5475\AA,
6425\AA~and 8075\AA, respectively.} (hereafter $VRI$) photometry for
all 104 objects to improve our distance estimates. These new distance
estimates are dubbed ``CCD distance estimates'' to reflect the
combination of CCD photometry at $VRI$ and 2MASS photometry at $JHK$,
and all $R$ and $I$ magnitudes refer to CCD values for the remainder
of this paper.

Care is taken during a night of photometry to ensure that previously
observed objects are sandwiched between targets that are new to the
program. Multiple results for a single object are then compared to
ensure consistency and can also be used to indicate possible
variability. The photometric observations used in making the distance
estimates reported here span a seven year interval from 2003 January
through 2010 May.\footnote{On the CTIO 0.9m, the Tek \#2 $VRI$ filter
set was used; however, the Tek \#2 $V$ filter originally used cracked
in March 2005 and was replaced by the very similar Tek \#1 $V$
filter. Reductions indicate no significant differences in the
photometric results from the two filters, as discussed in detail in
Jao et al.~2010. On the CTIO 1.0m, the Y4KCam filter set was used that
contains the same filters as those on the 0.9m.}  In total, over 800
frames were acquired on the 0.9m telescope at the Cerro Tololo
Inter-American Observatory (CTIO) for this study.  A small subset of
sixteen frames was acquired on the CTIO 1.0m in 2005 August; results
are consistent to 0.03 mag with the 0.9m data.

All data were reduced using IRAF. Calibration frames taken at the
beginning of each night were used for typical bias subtraction and
dome flat-fielding. Standard star fields from Graham (1982), Bessel
(1990) and/or Landolt (1992, 2007) were observed multiple times each
night in order to derive transformation equations and extinction
curves. In order to match those used by Landolt, apertures 14\arcsec
~in diameter were used to determine the stellar fluxes, except in
cases where close contaminating sources needed to be deconvolved. In
these cases, smaller apertures were used, and aperture corrections
were applied. Further details about the data reduction procedures,
transformation equations, etc., can be found in Jao et al.~(2005).

Figures 2$-$4 illustrate the total errors vs.~magnitude in each of the
three filters for the 268 stars with three or more nights of
photometry from our program, including some of the SCR stars reported
here.  These errors incorporate signal-to-noise errors, nightly errors
determined by fits to standard stars, and night-to-night measurement
differences to provide a comprehensive assessment of our photometric
errors. Signal-to-noise errors are 0.05 mag or less for objects
brighter than 16$^{th}$ mag at $V$, brighter than 18$^{th}$ mag at
$R$, and brighter than 17$^{th}$ mag at $I$. Standard star fits on
most nights have errors of 0.01$-$0.02 mag.  Typical variations in
photometric measurements from night to night are 0.02$-$0.03 mag.  As
can be seen in Figures 2$-$4, the convolution of all three errors
typically results in total errors of 0.03 mag or less for 66\% of
objects at $V$ and 84\% of objects at both $R$ and $I$, as evident
from the inset histograms. Note that the lack of a general increase in
errors for fainter stars is caused by observers increasing integration
times to provide adequate signal-to-noise on even the faintest
targets.  Nonetheless, these red dwarfs are faintest at $V$, so the
resulting errors are somewhat larger than at $R$ and $I$.  At $V$,
94\% of the objects presented have total errors of 0.06 mag or less;
for $R$ and $I$, this statistic is 99\%.  The SCR systems reported
here follow the general trend reported above: the fraction of objects
with errors less than 0.03 mag are 73\%, 94\% and 92\%, respectively,
at $VRI$.  At $V$, 94\% of the objects have total errors less than
0.06 mag, while this is true of 100\% of the objects at $R$ and $I$.

\subsection{$JHK$ Photometry from 2MASS}

Infrared photometry in the $JHK_{s}$ system has been extracted from
2MASS and is rounded to the nearest hundredth mag. The same 2MASS
photometry has been used for both the plate distance estimates and new
CCD distance estimates presented here.  The $JHK_{s}$ magnitude errors
listed are the x$_{-}$sigcom errors (where x is j, h, or k) and
include target, global, and systematic influences. These errors are
generally less than 0.04 mag. Exceptions at $H$ are SCR0214-0733 (0.05
mag), SCR0211-6108, SCR1230-3411, and SCR0717-0500 (all 0.06 mag).

\subsection{CCD Distance Estimates}
\label{sec:phot.ccd}

The CCD distance estimates\footnote{The term 'CCD distance estimate'
is used for the remainder of the paper to indicate a CCD photometric
distance estimate.}, as described in Henry et al.~(2004), are found
using a method similar to that used for the photographic plate
distance estimates. The difference is that we use the accurate $VRI$
magnitudes obtained at CTIO instead of $BRI$ plate magnitudes from
SuperCOSMOS. The maximum number of relations possible from the
combination of $VRIJHK$ magnitudes is 15 (the same number of
combinations as for the plate relations), but only 12 yield useful
results. The color spread is not large enough in three of the
relations ($M_K$ vs.~$J-H$, $J-K$, and $H-K$) to give reliable
results and so are omitted.

Of the 100 systems with photographic plate distance estimates that
place them closer than 25.0 parsecs, we find 77 to have CCD distance
estimates within the 25.0 pc horizon. Thus, we retain 77\% of the
SuperCOSMOS discoveries as likely nearby stars. All systems with CCD
distance estimates within 15.0 pc and additional compelling systems
are targeted for trigonometric parallax measurements as part of our
Cerro Tololo Inter-American Observatory Parallax Investigation
(CTIOPI) (Jao et al.~2005, Henry et al.~2006, Subasavage et al.~2009,
Riedel et al.~2010). At present, 31 of the systems discussed here are
on the CTIOPI program.

Figure 5 shows a comparison of the two photometric distance estimates
with error bars omitted for clarity. Primary stars and a few
secondaries for which separate photometry is available are
plotted. The single point in the upper right of the plot and one point
with a plate distance estimate of 25--30 pc are secondary components
(SCR2241-6119B and SCR2335-6433B). Twenty-three systems are estimated
to be beyond 25 pc using CCD distance estimates. The two most likely
causes of the distance offsets are poor plate photometry compared to
CCD photometry and close multiples that are unresolved on the plates
but resolved in CCD images. Errors in the plate photometry certainly
cause most of the discrepancies. A distance of up to $\sim$35 pc is
possible if the object is an unresolved binary in which the two
components are identical and contribute an equal amount of light to
the measured magnitude. All SCR targets have been inspected for
duplicity in the CCD images. Seven were found to be binaries with
separations between 1\arcsec~(typical resolution of our CCD images)
and 5\arcsec. The typical resolution of the plate images for binaries
with magnitude differences of $\lesssim$ 2 is 4\arcsec. Details on
multiple systems can be found in \S 5.1.

Eight objects are found to have CCD distance estimates of 35$-$65 pc.
One object (SCR2335-6433B in the upper right of Figures 5 and 6) is a
common proper motion companion that was discovered by eye when
blinking plates. The remaining seven objects had 3$-$8 plate relations
contributing to their distance estimates rather than the full suite of
11, indicative of inaccurate estimates.  The corresponding CCD
distances for these objects often rely on 6$-$10 relations rather than
the full suite of 12, typically because they are bluer than some of
the plate and CCD distance relations.  Three of the eight objects have
already been verified to be K dwarfs via spectroscopy.

\subsection{Photometric Distance Estimate Errors}
\label{sec:phot.errors}

When using both the plate and CCD distance estimating suites, there
is inherent spread around the fit line for each relation that is a
result of two causes: (1) the deviations due to errors in parallax and
photometry for the stars used to derive the fits, and (2) cosmic
scatter due to differences in stellar ages, compositions, and perhaps
magnetic properties.  Cosmic scatter dominates the offsets between
distance estimates and true distances. Unresolved, unknown multiples
may also contribute, although great care was taken to remove such
systems before the fits were made, so this is not a significant issue.
In order to estimate the reliability of the suites of relations in
both the plate and CCD methods, we have run single, main sequence, red
dwarfs with known trigonometric distances back through the relations
to derive representative errors. The median offsets between the
trigonometric and photometric distance estimate are 26\% for the plate
suite and 15\% for the CCD suite.

For all errors on distance estimates listed in Table 3, these
percentage errors are combined in quadrature with each star's
individual error, which comes from the up to 11 or 12 distances
derived from the plate and CCD suites, respectively. Figure 6
is the same as Figure 5, but with the total errors shown for both
estimates. Note that in no case can the plate distance estimate have
an error less than 26\% nor less than 15\% for CCD distance estimate.

Figure 7 illustrates a comparison of the SuperCOSMOS plate ($R-I$)
color and the CCD ($R-I$) color for 100 SCR stars reported here.  The
fit is a polynomial described by

($R-I$)$_{ccd}$ $=$ 0.640 + 0.192*($R-I$)$_{plate}$ +
0.198*($R-I$)$^2$$_{plate}$ - 0.035*($R-I$)$^3$$_{plate}$.

This polynomial permits users of SuperCOSMOS data to predict the
$(R-I)$$_{ccd}$ color that would be measured for red objects,
i.e. those with $-$0.2 $=$ ($R-I$)$_{plate}$ $=$ 3.8.

Two outlying points represented by open squares are SCR1448-5735 and
SCR1637-4703, both of which were omitted from the fit because they are
blended with other sources in $R$ images.  Two other objects for which
$I$ plate magnitudes were not available (SCR1931-0306 and
SCR2230-5244), were also excluded from the fit. The mean absolute
deviation is 0.27 mags, indicating that the SuperCOSMOS $(R-I)$ color
is roughly nine times less accurate than the CCD $(R-I)$ color, which
typically has errors less than 0.03 mag.

\section{Confirmation as Red Dwarfs: Spectroscopy and the Color-Color Diagram}
\label{sec:spec}

For 37 stars, we have spectra that were acquired on the CTIO 1.5m
between 2003 July and 2006 December. Observations were made using a
2\farcs0~slit or wider in the RC Spectrograph with grating \#32 and
order blocking filter OG570 to provide wavelength coverage from 6000
to 9500 \AA~and resolution of 8.6 \AA~on the Loral 1200x800
CCD. Reductions were done using standard procedures via IRAF.
Spectral types have been assigned on the RECONS system (Jao et
al.~2008).  Thirty-four of the stars are confirmed to be M dwarfs,
while three are K dwarfs.  The three K dwarfs have slipped into the
sample because of erroneous plate magnitudes, as evidenced by having
fewer than 11 relations for their plate distance estimates.

For all of the stars, we use a color-color diagram to verify that the
sample objects are red dwarfs.  Figure 8 is a color-color diagram that
plots ($V-K$) versus ($J-K$). Labels denoting spectral types are shown
inside the left axis. SCR stars are open circles, with the combined
photometry of seven SCR close binaries marked by crosses inside the
circles. The dots are RECONS 10 pc sample members (Henry et al.~2006),
and a few known giants are plotted as solid triangles for
reference. As is evident from this diagram, none of the SCR stars
presented in this paper fall in the region where known giants are
found, so we are relatively confident that none of these objects are
giants. We anticipate that these objects are not subdwarfs because
their tangential velocities are $\leq$ 100 km/s. The point that is
slightly elevated (1.02, 5.90) toward the three giants is the binary
system SCR1630-3633AB.

\subsection{Multiples}
\label{sec:disc.multiples}

Eleven of the systems presented in this paper are binary, with seven
systems having separations less than 3\arcsec. Six of these close
binaries are newly discovered via the SCR effort --- SCR0630-7643AB,
SCR0644-4223AB, SCR1630-3633AB, SCR1856-4704AB, SCR1926-8216AB, and
SCR2042-5737AB --- while SCR1845-6357A was discovered via the SCR
effort and the companion first detected by Biller et al.~(2006).
Table 4 lists the binary data, including the separations of the
components, position angles of the secondaries relative to the
primaries, distance estimates of the system for close systems
(indicated by brackets) and of the components where separate
photometry was possible, and the distance errors.

All systems were re-blinked using the SuperCOSMOS photographic plates
to ensure that no widely separated companions out to 5\arcmin~had been
previously overlooked. No new discoveries were made. A search was then
made for closer companions by examining all $I$ frames taken at CTIO
in IRAF, using the surface, contour and radial plot features
available. Separations were determined using the SExtractor (Bertin \&
Arnouts 1996) feature in IRAF, with some parameters ({\it
detect\_minarea, deblend\_nthresh, deblend\_mincont}) modified in
order for the program to recognize both components of the close
system. These results are listed in Table 4.

Two binary systems, SCR0630-7643AB and SCR1630-3633AB, have been
confirmed to be physically related via HST-FGS (Hubble Space Telescope
Fine Guidance Sensors) observations.  The $\Delta$$V$ values between
the two components in each system were estimated by comparing HST-FGS
scans to systems in which the $\Delta$$V$ is known (Henry et
al.~1999).

\subsection{Objects Worthy of Note}
\label{sec:disc.weirdos}

{\bf SCR0630-7643AB} has a photometric distance estimate of 5.5 pc,
but has a trigonometric distance of 8.8 pc in Henry et al.~2006. This
is a new close binary with a separation of 1\farcs0 and a
$\Delta$$V$~$\thicksim$~0.3 mag that has been confirmed with both
continuing optical speckle and HST-FGS observations.

{\bf SCR1107-3420B} is 30$\farcs$6 from the primary at a position
angle of 107.1$^\circ$; the primary component is a white dwarf with an
estimated distance of 28.2 pc (Subasavage et al.~2007). No companions
have been found for the red dwarf secondary by HST-FGS and speckle
observations.

{\bf SCR1630-3633AB} is a new binary system, confirmed by HST-FGS with
$\Delta$$V$~$\thicksim$~0.7 mag.  A more realistic distance estimate is
$\sim$20 pc, taking into consideration the system's multiplicity.

{\bf SCR1746-8211} was thought to be a possible common proper motion
companion to HD 158866, at a separation of 76$\farcs$5 at
290.8$^{\circ}$ (Finch et al.~2007).  However, the distance estimate,
15.5 pc, is not in agreement with the distance of the primary, 30.6
pc, as measured by HIPPARCOS (van Leeuwen 2007). The proper motion of
the primary, $\mu=$ 0\farcs211/yr at 191$^\circ$, is inconsistent with
that of the secondary, $\mu=$ 0\farcs288/yr at 184.9$^{\circ}$.  Thus,
the two objects do not appear to be in the same system.

{\bf SCR1845-6357AB} has a CCD distance estimate of 4.6 pc and a
trigonometric parallax that revises the distance to 3.9 pc (Henry et
al.~2006). This is the 24$^{th}$ closest system to the Sun, composed
of M8.5 and T dwarfs at a separation of 1$\farcs$2 (Biller et
al.~2006).

{\bf SCR2241-6119AB} is a binary system separated by 9.6$\arcsec$ at
211.2$^{\circ}$.  The two components appear to be co-moving in
SuperCOSMOS images, but the proper motion values do not match: A has
$\mu$ $=$ 0$\farcs$184/yr at 124.0$^{\circ}$ while B has $\mu$ $=$
0$\farcs$287/yr at 108.0$^{\circ}$.  However, the B component is very
faint in the $B$ image, faint in the $R$ image, and not round in $I$
image, likely leading to a poor, and inconsistent, proper motion
measurement.
 
{\bf SCR2335-6433AB} has a separation of 22$\farcs$4 at 25.9$^\circ$
(Finch et al.~2007), making photometry for each component possible.
However, the distance estimates for the A and B components, 35.9 and
56.7 pc, respectively, do not agree. This is likely due to the fact
that only nine of the 12 CCD relations were in agreement for the
primary. As all 12 relations were utilized in the distance estimate of
the secondary, the true distance of the system is probably closer to
that of the secondary.

\section{Discussion}
\label{sec:discussion}

The majority of the objects (71 of 104, or 68\% of the sample) have
$V>$14 because these new discoveries are fainter than those targeted
by Luyten in most of the southern sky.  Only the faint companion
object discovered by eye during the blinking process, SCR2241-6119B
($R =$ 17.23), is fainter than CCD $R =$ 16, a consequence of the
plate $R =$ 16.5 cut-off chosen in our searches to date. The color-color
diagram in Figure 8 shows that most of the 104 SCR objects are M3$-$M6
type dwarfs, which corresponds to the ``sweet spot'' created by our
search magnitudes through most of the 25 pc volume.

As can be seen in the polar plot of Figure 1, many of these SCR
discoveries are south of $-$60$^\circ$. Luyten's Palomar survey for
proper motion stars extended to $-$45$^\circ$ for a quarter of the
southern sky and to $-$33$^\circ$ for the entire southern sky (Luyten
1974), and his Bruce Survey covered the remainder of the southern sky.
Giclas searched declinations north of $-$45$^\circ$ for stars of
magnitudes 8$-$17 with proper motions $\geq$~0.20 and at
even smaller $\mu$ for red objects (Giclas et al.~1978a).  Still, we
have found 11 systems between DEC $=$ 0$^\circ$ and $-$30$^\circ$, an
area of the sky covered previously by both Luyten and Giclas. Four of
these eleven are brighter than $V=$14 and so were missed by those
surveys.

In total, the 77 systems we determine to be within 25 pc, when
compared to the 329 southern systems found in NStars, result in a 23\%
increase in the number of red dwarf systems nearer than 25.0 pc in the
southern sky.  In Table 5 we summarize the distance distribution for
the 100 systems discussed in this paper, as well as the number of
binaries found within each distance bin.  We anticipate that seven
will prove to be within 10 pc and 70 will be between 10 and 25 pc,
while the remaining 23 systems are likely beyond 25 pc.  If all of the
new red dwarf systems from our SCR search to date, plus the 89 systems
found by Phan-Bao et al.~, Reyl{\'e} et al., Finch et al.,and Costa \&
M{\'e}ndez are found to have trigonometric parallaxes placing them
within 25 pc, the number of red dwarfs systems in the southern sky
within that horizon would increase by 50\%.  Forthcoming in this
series is a paper presenting trigonometric parallaxes for roughly
one-third of the SCR systems in this paper, three of which
(SCR0630-7643AB, SCR1138-7721, SCR1845-6357AB) have had results
previously published (Henry et al.~2006).

We have several continuing efforts to reveal the Sun's red dwarf
neighbors.  We have found another $\sim$100 systems with 0\farcs40/yr
$>$ $\mu$ $\ge$ 0\farcs18/yr between DEC $=$ 0$^\circ$ and
$-$47$^\circ$ during our continuing sweep of the southern sky.  This
search is a complement to our Finch et al.~2007 effort and will
complete the SCR search of the southern sky for stars that would have
been members of Luyten's New Luyten Two-Tenths (NLTT) Catalogue
(Luyten 1979b).  Already underway is a project to gather photometric
data for $\sim$150 additional SCR systems that have photographic plate
distance estimates placing them between 25 and 30 parsecs.  Some of
these systems will, in effect, replace the 23 systems with CCD
distances beyond the 25 pc original limit of the plate distances.
Already, of the thirty-eight systems for which we have obtained
initial $VRI$ photometry, nine are anticipated to be nearer than the
25 pc horizon.  Still remaining are additional searches that probe
beyond the $R = 16.5$ cutoff of our first sweep of the southern sky.

\section{Acknowledgments}

The support of additional RECONS team members was vital for this
research, in particular Adric Riedel.  We also thank Brian Mason for
insight on preliminary optical speckle survey results, as well as Ed
Nelan for early results from HST-FGS.  This work has been supported
under National Science Foundation grants AST 05-07711 and AST
09-08402, as well as by NASA's Space Interferometry Mission, and
Georgia State University.  We also thank the members of the SMARTS
Consortium, who enable the operations of the small telescopes at CTIO,
as well as the observer support at CTIO, especially Edgardo Cosgrove,
Arturo Gomez, Alberto Miranda, and Joselino Vasquez. This research
makes use of data products from the Two Micron All Sky Survey, which
is a joint project of the University of Massachusetts and the Infrared
Processing and Analysis Center/California Institute of Technology,
funded by the National Aeronautics and Space Administration and the
National Science Foundation.  This research also makes use of the
SIMBAD database and the Aladin and Vizier interfaces, operated at CDS,
Strasbourg, France. Thank you to the referee for their helpful
comments, as well.


\clearpage


\begin{figure}
\includegraphics[scale=0.90,angle=90]{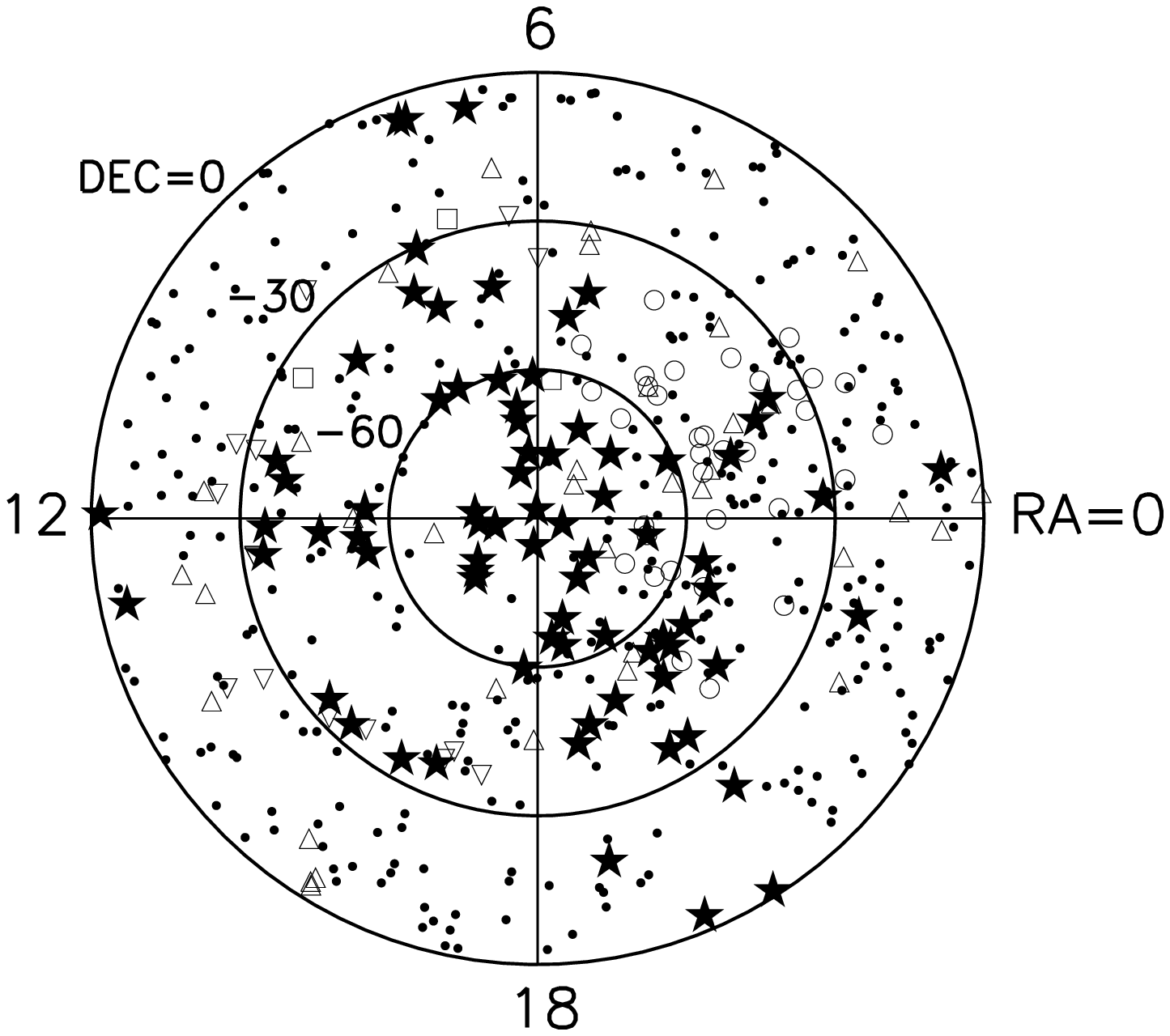}
 
\figcaption[fig1] {Polar plot for southern hemisphere photometric
distance estimates of objects within 25.0 pc, where RA advances
counterclockwise from zero to 24 hours, and DEC ranges from 0$^\circ$
at the outer edge to $-$90$^\circ$ in the center. The 329 red dwarf
primaries with trigonometric parallaxes from NStars are noted as small
solid black dots; Phan-Bao et al.~results are shown as 37 open
triangles; Reyl\'e et al.~results are shown as 34 open circles; Finch
et al.~results are 15 open upside-down triangles; Costa \& M{\'e}ndez
results are 3 open squares; RECONS' SCR results are shown as 77 solid
stars from this paper with CCD photometric distance
estimates. \label{fig:winters1}}

\end{figure}
 
\clearpage
 
\begin{figure}
\includegraphics[scale=0.60,angle=270]{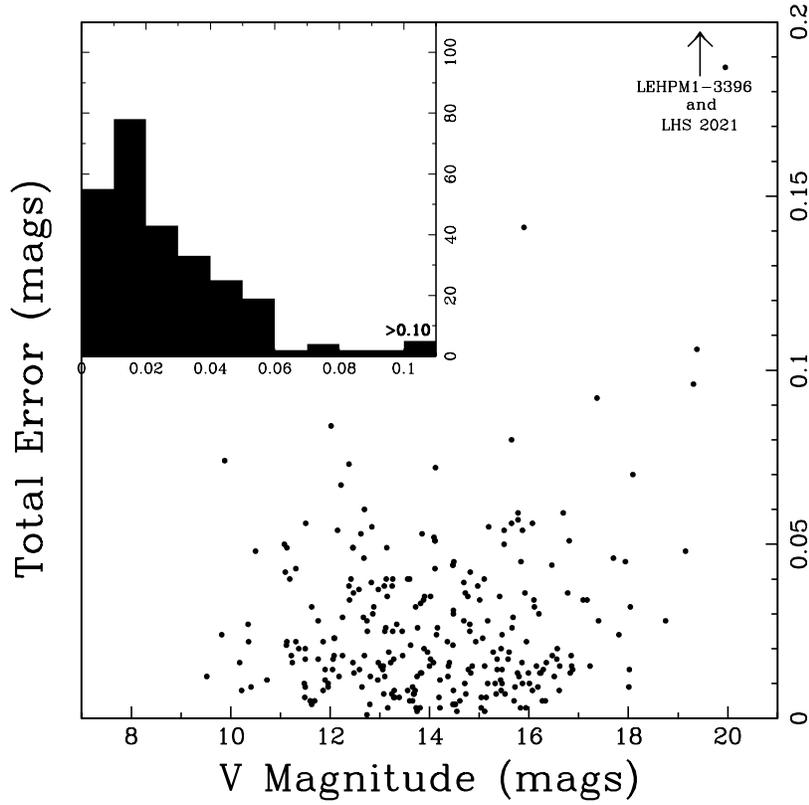}

\figcaption[fig2] {Total CCD photometric errors vs.~$V$ magnitude for
the CTIO 0.9m telescope data. Inset at the upper left is a histogram
of the number of objects in each error bin. The total number of
objects plotted is 268 (including some of the SCR objects presented in
this paper), and each error bin is 0.01 mag. The arrow at the upper
right of the figure indicates that LEHPM1-3396 (V$=$19.38) and LHS 2021
(V$=$19.21) are both off the plot. \label{fig:winters2}}

\end{figure}
 
\clearpage

\begin{figure}
\includegraphics[scale=0.60,angle=270]{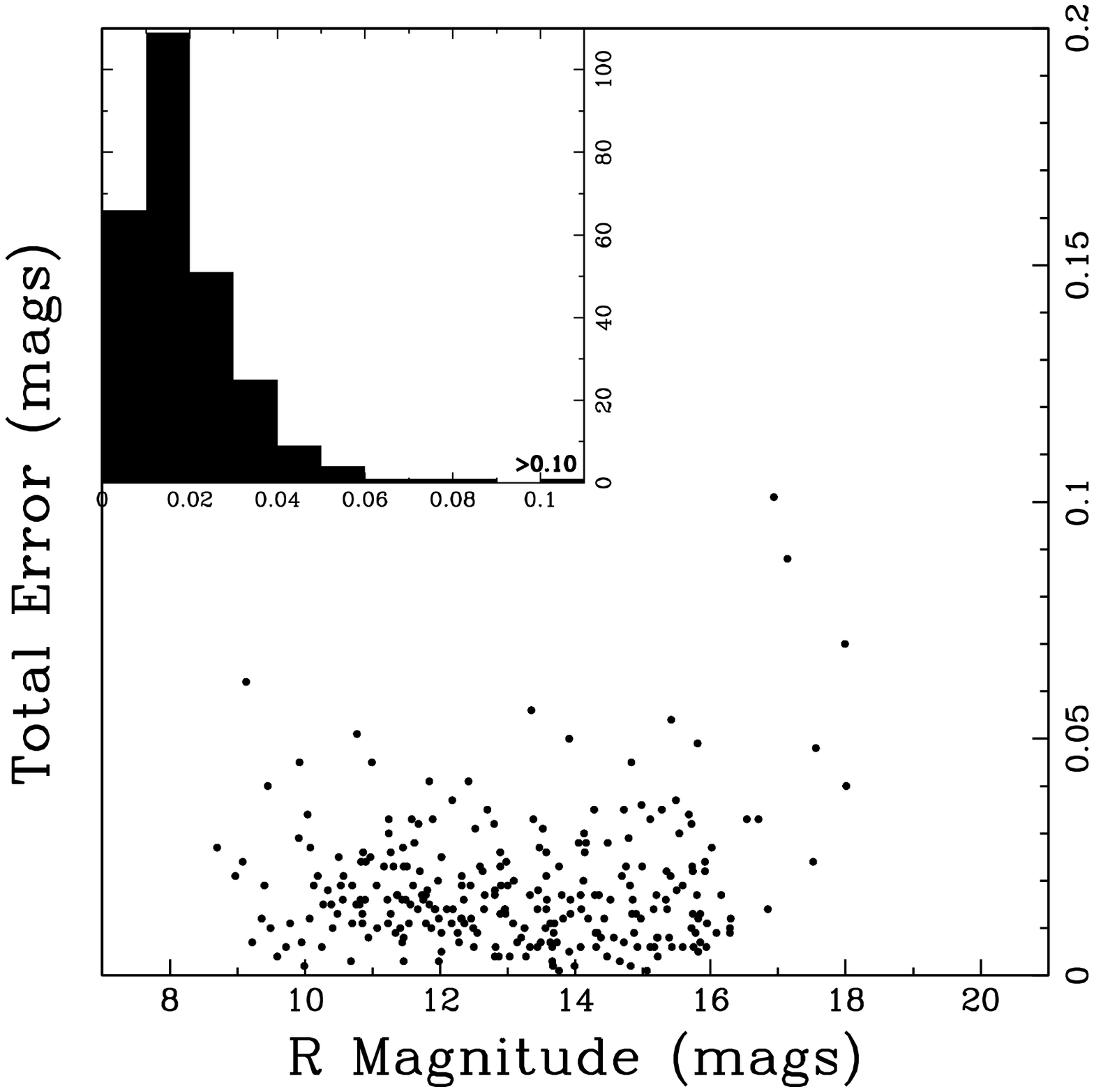}

\figcaption[fig3] {Same as Figure 2, but for CCD photometry in the $R$
band. \label{fig:winters3}}

\end{figure}
 
\clearpage
 
\begin{figure}
\includegraphics[scale=0.60,angle=270]{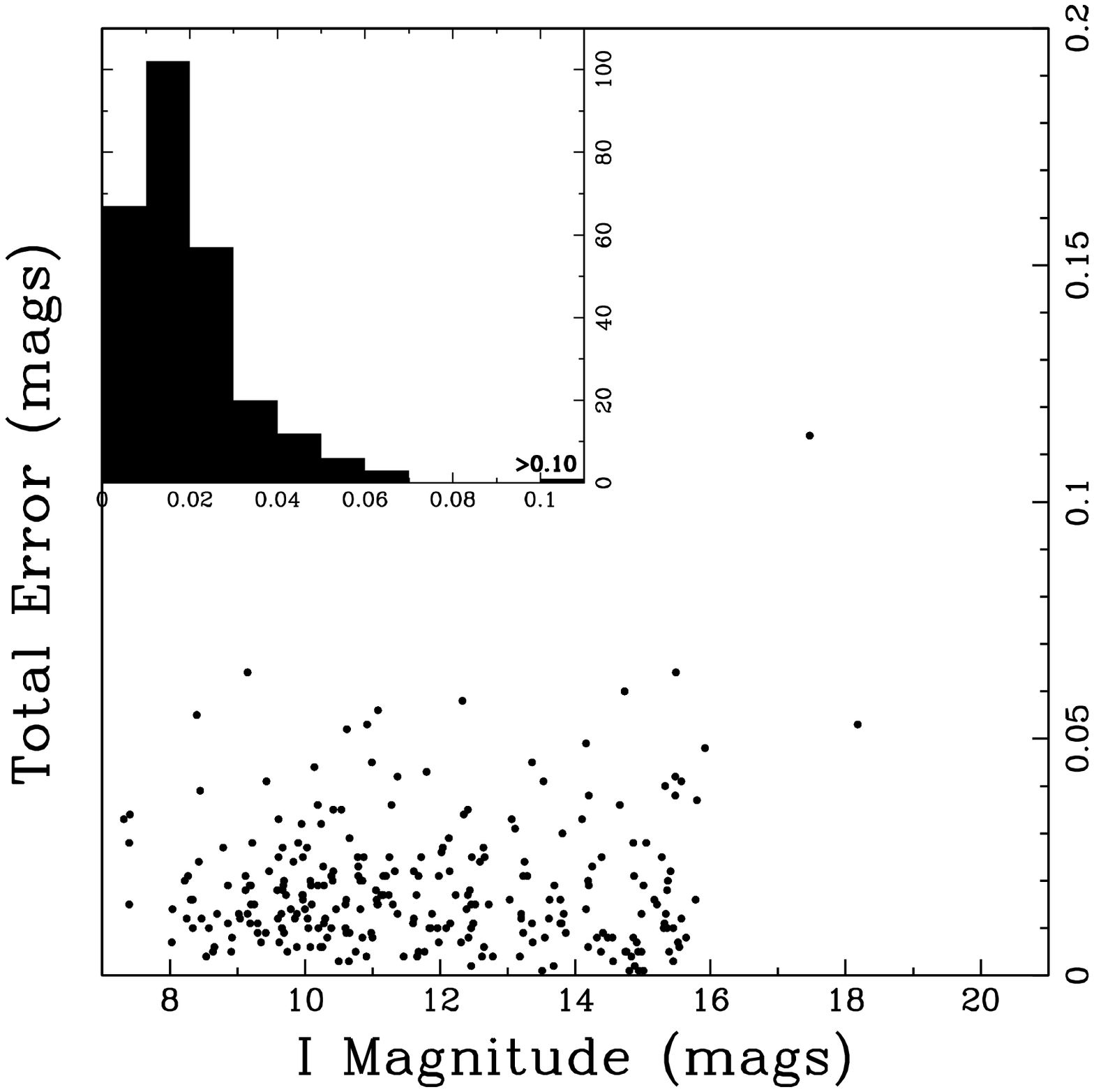}

\figcaption[fig4] {Same as Figure 2, but for CCD photometry in the $I$
band. \label{fig:winters4}}

\end{figure}
 
\clearpage
 
\begin{figure}
\includegraphics[scale=0.60,angle=270]{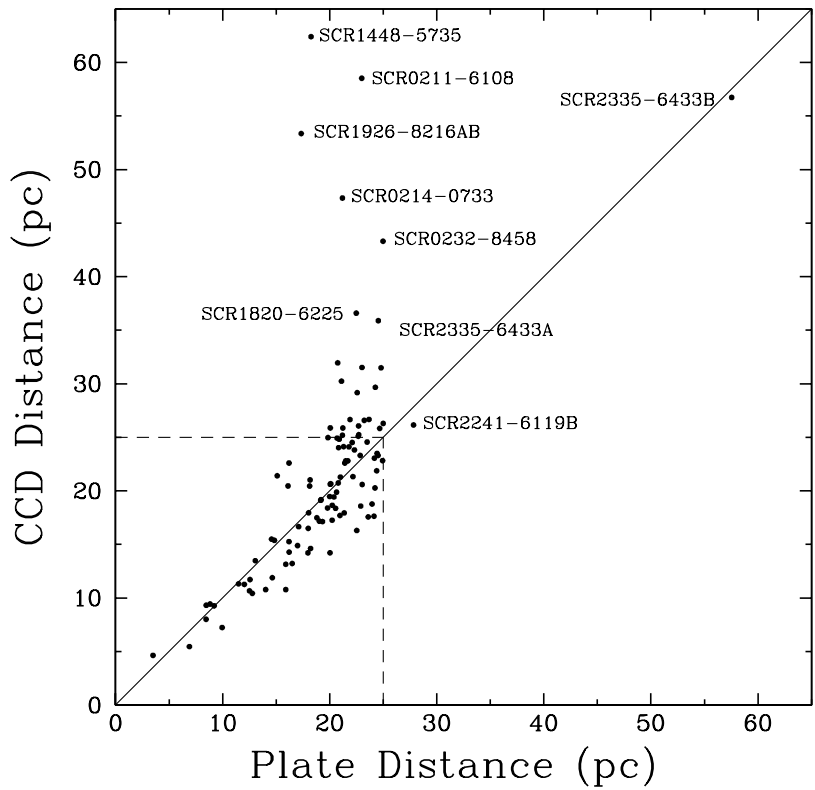}

\figcaption[fig5] {CCD distance vs. plate distance without error
bars. A solid diagonal line represents CCD distance estimates exactly
matching photographic plate distance estimates. The dashed line is a
25.0 pc boundary, indicating the ultimate distance horizon of interest
for this sample. Outliers have been labeled. \label{fig:winters5}}

\end{figure}
 
\clearpage
 
\begin{figure}
\includegraphics[scale=0.60,angle=270]{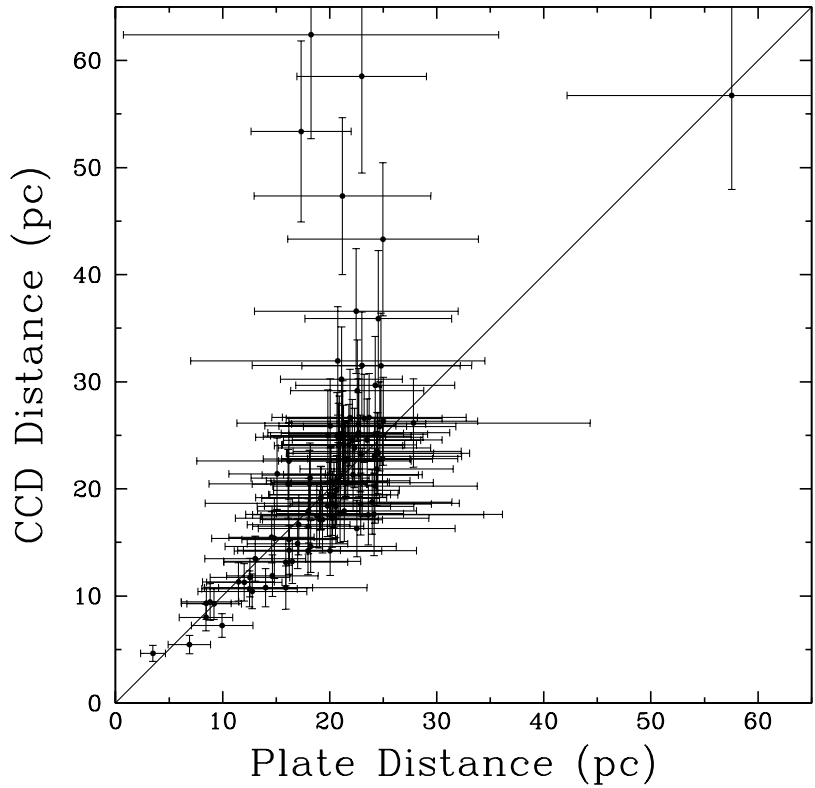}

\figcaption[fig6] {CCD distance vs. plate distance with error bars.
The total errors for photographic plate distance estimates include the
standard deviation of the (up to) 11 individual distance estimates
plus 26\% systematic errors incorporated primarily because of cosmic
scatter.  Similarly, the total errors for CCD distance estimates
include the standard deviation of the (up to) 12 individual distance
estimates plus 15\% systematic errors incorporated primarily because
of cosmic scatter. \label{fig:winters6}}

\end{figure}
 
\clearpage
 
\begin{figure}
\includegraphics[scale=0.80,angle=0]{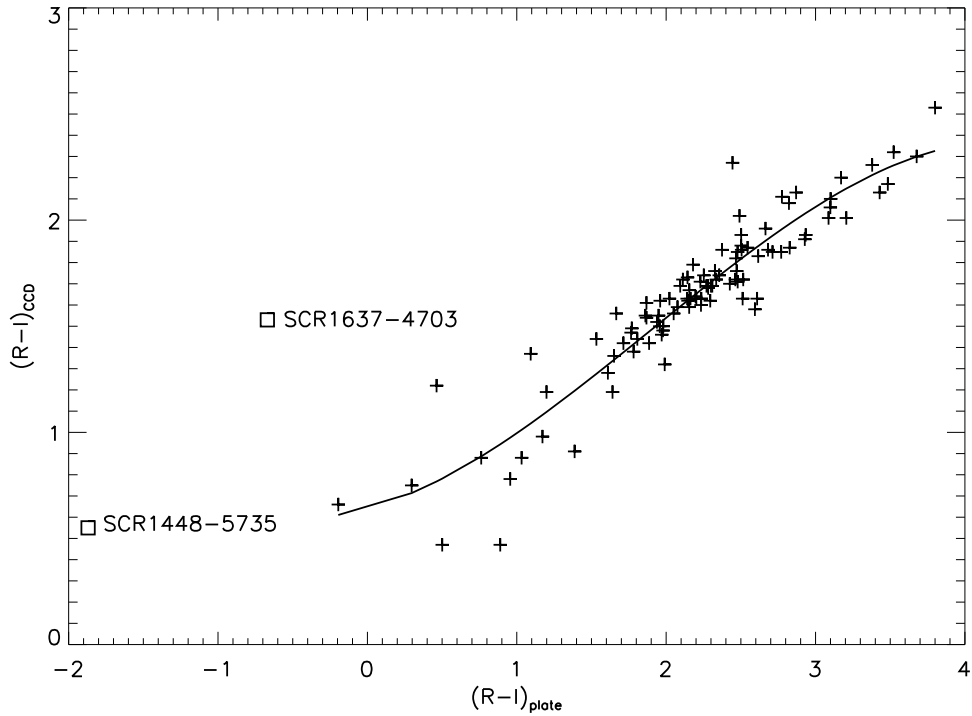}

\figcaption[fig7] {CCD ($R-I$) color vs. Plate ($R-I$) color with a
polynomial fit for 100 SCR objects. The two outliers, denoted by open
squares and labeled, were not used in the fit due to blending on the
$R$ plates, nor were two objects for which no $I$ plate magnitude was
available (SCR1931-0306 and SCR2230-5244).\label{fig:winters7}}

\end{figure}
 
\clearpage
\begin{figure}
\includegraphics[scale=0.60,angle=270]{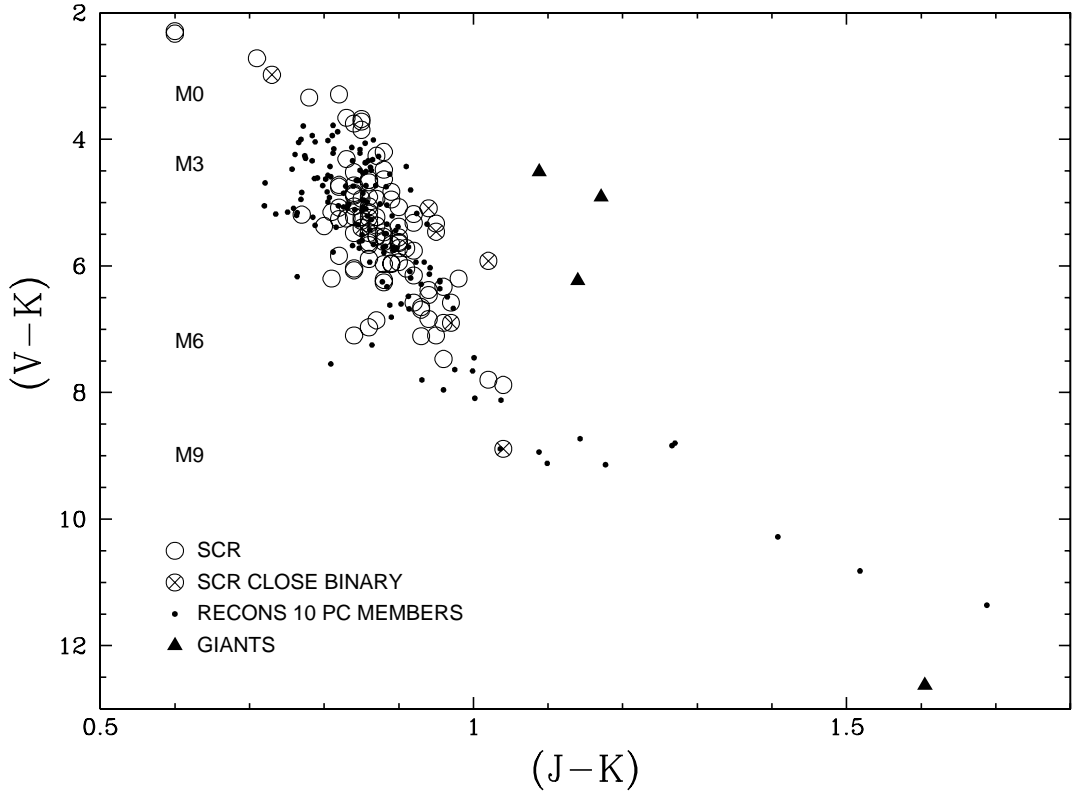}

\figcaption[fig8] {Color-color diagram of $(V-K)$ versus
$(J-K)$. Spectral type estimates are listed inside the y-axis. Open
circles indicate 104 SCR objects presented in this paper; for the two
resolved binary systems (SCR2241-6119AB and SCR2335-6433AB),
photometry is available for both components. X's surrounded by open
circles symbolize confirmed SCR binaries with combined
photometry. Known RECONS 10 pc sample systems are shown as solid dots,
and known giants are denoted by solid triangles. \label{fig:winters8}}

\end{figure}
 
\clearpage


\voffset0pt{ \centering
\begin{deluxetable}{lcc}
\tabletypesize{\scriptsize}
\tablecaption{Recent Photometric Distance Estimation Efforts in the Southern Sky}
\setlength{\tabcolsep}{0.03in}
\label{tab:others.photpi}
\tablewidth{0pt}

\tablehead{\colhead{Source}                          &
 	   \colhead{\# Systems}                        &
 	   \colhead{Reference}                       }
                                                        
\startdata
NStars                      &      329\tablenotemark{a} &  1,2\\
\hline
Phan-Bao et al.             &       37      &  3,4,5,6,7\\
Reyl{\'e} et al.            &       34      &  8,9\\ 
Finch et al.                &       15      &  10\\
Costa \& M{\'e}ndez         &        3      &  11\\ 
{\bf RECONS (plate phot)}   & {~\bf 100}    &  12,13,14,15,16,*\\
{\bf RECONS (CCD phot)}     & {~\bf 77}     &  *\\
\enddata

\tablenotetext {a} {trigonometric parallaxes $\geq$ 40 milliarcseconds}

\tablecomments{
(1): {NStars Database, see Henry et al.~2002}; 
(2): {NStars Database, see Henry et al.~2003}; 
(3): {Delfosse et al.~2001}; 
(4): {Phan-Bao et al.~2001}; 
(5): {Phan-Bao et al.~2003}; 
(6): {Phan-Bao et al.~2006}; 
(7): {Phan-Bao et al.~2008}; 
(8): {Reyl{\'e} et al.~2002}; 
(9): {Reyl{\'e} \& Robin(2004)};
(10): {Finch et al.~2010};
(11): {Costa \& M{\'e}ndez 2003}; 
(12): {Hambly et al.~2004};
(13): {Henry et al.~2004}; 
(14): {Subasavage et al.~2005a}; 
(15): {Subasavage et al.~2005b}; 
(16): {Finch et al.~2007};
(*): {this paper}.}

\end{deluxetable}
}


\voffset000pt{
\centering
\begin{deluxetable}{lccrrr}
\setlength{\tabcolsep}{0.03in}
\label{tab:bri}
\tablewidth{0pt}
\tabletypesize{\scriptsize}
\tablecaption{$BRI$ Photometric Results for New SCR Discoveries}
\tablehead{\colhead{Name}                &
	   \colhead{R.A.}                &
 	   \colhead{DEC}               &
           \colhead{$B_J$}               &
           \colhead{$R_{59F}$}           &
           \colhead{$I_{IVN}$}           }

\startdata
SCR0017-3219     & 00 17 15.73 & $-$32 19 54.0 & 16.80  & 14.59  & 12.48  \\      
SCR0027-0806     & 00 27 45.36 & $-$08 06 04.7 & 18.65  & 16.26  & 13.49  \\      
SCR0113-7603     & 01 13 31.47 & $-$76 03 09.3 & 17.68  & 15.23  & 13.05  \\      
SCR0137-4148     & 01 37 23.49 & $-$41 48 56.2 & 16.91  & 14.55  & 12.24  \\      
SCR0150-3741     & 01 50 13.25 & $-$37 41 52.0 & 15.80  & 13.49  & 11.41  \\      
SCR0214-0733     & 02 14 44.69 & $-$07 33 24.7 & .....  &  9.71  &  8.82  \\
SCR0325-0308     & 03 25 03.09 & $-$03 08 20.4 & 15.22  & 13.17  & 11.39  \\      
SCR0506-4712     & 05 06 07.29 & $-$47 12 51.6 & 14.07  & 11.96  & 10.32  \\      
SCR0509-4325     & 05 09 43.85 & $-$43 25 17.4 & 15.12  & 13.00  & 10.71  \\      
SCR0513-7653     & 05 13 05.97 & $-$76 53 21.9 & 14.54  & 12.32  & 11.23  \\      
SCR0526-4851     & 05 26 40.80 & $-$48 51 47.3 & 13.82  & 11.67  & 10.02  \\      
SCR0607-6115     & 06 07 58.09 & $-$61 15 10.5 & 15.75  & 13.43  & 11.13  \\      
SCR0643-7003     & 06 43 29.79 & $-$70 03 20.8 & 13.95  & 11.51  &  9.75  \\      
SCR0644-4223AB   & 06 44 32.09 & $-$42 23 45.2 & 15.59J & 13.35J & 11.18J \\      
SCR0713-0511     & 07 13 11.23 & $-$05 11 48.6 & 11.76  &  9.31  &  8.84  \\
SCR1107-3420B    & 11 07 50.25 & $-$34 21 00.6 & 16.35  & 14.15  & 11.81  \\
SCR1932-0652     & 19 32 46.33 & $-$06 52 18.1 & 14.92  & 13.27  & 11.50  \\
SCR2009-0113     & 20 09 18.24 & $-$01 13 38.2 & 15.36  & 13.09  & 10.62  \\
SCR2018-3635     & 20 18 06.52 & $-$36 35 27.7 & 16.31  & 14.07  & 11.47  \\      
SCR2025-2259     & 20 25 18.93 & $-$22 59 06.0 & 15.26  & 13.26  & 11.40  \\      
SCR2033-4903     & 20 33 01.87 & $-$49 03 10.6 & 16.57  & 14.17  & 11.34  \\      
SCR2104-5248     & 21 04 53.85 & $-$52 48 34.3 & 14.72  & 12.75  & 10.94  \\      
SCR2105-5503     & 21 05 13.78 & $-$55 03 56.3 & 15.01  & 12.96  & 10.76  \\      
SCR2122-4314     & 21 22 16.92 & $-$43 14 05.0 & 14.26  & 12.06  & 10.01  \\      
SCR2135-5325     & 21 35 39.63 & $-$53 25 31.5 & 17.43  & 15.07  & 12.30  \\      
SCR2142-7405     & 21 42 58.54 & $-$74 05 55.5 & 14.24  & 12.20  & 10.59  \\      
SCR2252-2220     & 22 52 25.82 & $-$22 20 06.8 & 14.95  & 12.82  & 10.85  \\      
SCR2301-5530     & 23 01 32.51 & $-$55 30 17.6 & 14.06  & 12.24  & 10.25  \\      
SCR2325-6740     & 23 25 25.13 & $-$67 40 07.9 & 15.60  & 13.32  & 11.09  \\    

\enddata

\tablecomments{coordinates are epoch and equinox 2000.0 using
SuperCOSMOS positions transformed to epoch 2000.0 using the proper
motions and position angles listed here; if AB, photometry or spectral
type is for the system, i.e. joint. In the case of SCR0214-0733, no
$B$ plate magnitude was available in SuperCOSMOS.}

\end{deluxetable}
}


\voffset000pt{
\centering
\begin{deluxetable}{lccccrrrcrrrrrcrrccccl}
\rotate
\setlength{\tabcolsep}{0.03in}
\label{tab:phot}
\tablewidth{0pt}
\tabletypesize{\scriptsize}
\tablecaption{Photometric Results for SCR Sample}
\tablehead{\colhead{Name}                &
	   \colhead{R.A.}                &
 	   \colhead{DEC.}               &
	   \colhead{$\mu$}               &
	   \colhead{P.A.}                &
           \colhead{$V_{J}$}             &
           \colhead{$R_{KC}$}            &
           \colhead{$I_{KC}$}            & 
           \colhead{\# nts}              &
           \colhead{$J$}                 &
           \colhead{$H$}                 &
           \colhead{$K_{s}$}             &
	   \colhead{d$_{plt}$}           &
           \colhead{$\sigma$$_{plt}$}    &
	   \colhead{\# rel}              &
	   \colhead{d$_{ccd}$}           &
           \colhead{$\sigma$$_{ccd}$}    &
	   \colhead{\# rel}              &
	   \colhead{$V_{tan}$}           &
	   \colhead{SpType}              &
	   \colhead{ref}                 &
	   \colhead{notes}               \\

	   \colhead{   }                 &
	   \colhead{   }                 &
           \colhead{   }                 &
           \colhead{(\arcsec/yr)}        &
           \colhead{deg}                 &
	   \colhead{   }                 &
           \colhead{   }                 &
           \colhead{   }                 &
           \colhead{   }                 &
           \colhead{    }                &
           \colhead{    }                &
           \colhead{    }                &
           \colhead{(pc)}                &
           \colhead{(pc)}                &
	   \colhead{    }                &
           \colhead{(pc)}                &
           \colhead{(pc)}                &
	   \colhead{    }                &
	   \colhead{(km/s)}              &
	   \colhead{    }                &
	   \colhead{    }                &
           \colhead{    }                \\

	   \colhead{(1)}                 &
	   \colhead{(2)}                 &
           \colhead{(3)}                 &
           \colhead{(4)}                 &
           \colhead{(5)}                 &
	   \colhead{(6)}                 &
           \colhead{(7)}                 &
           \colhead{(8)}                 &
           \colhead{(9)}                 &
           \colhead{(10)}                &
           \colhead{(11)}                &
           \colhead{(12)}                &
           \colhead{(13)}                &
	   \colhead{(14)}                &
	   \colhead{(15)}                &
	   \colhead{(16)}                &
	   \colhead{(17)}                &
	   \colhead{(18)}                &
	   \colhead{(19)}                &
	   \colhead{(20)}                &
	   \colhead{(21)}                &
           \colhead{(22)}                }

\startdata
SCR0017-3219     & 00 17 15.73 & $-$32 19 54.0 & 0.220 & 096.9 & 15.45  & 14.11  & 12.39  & 2 & 10.64  & 10.08  &  9.73  ~& 21.0 &  7.3 & 11 & 21.3 & 3.4 & 12 & 22.2 & ----- &   6   & \\      
SCR0027-0806     & 00 27 45.36 & $-$08 06 04.7 & 0.184 & 122.3 & 17.51  & 15.83  & 13.72  & 2 & 11.57  & 10.97  & 10.61  ~& 22.9 &  6.6 & 11 & 18.6 & 2.9 & 12 & 16.2 & ----- &   6   & \\      
SCR0111-4908     & 01 11 47.52 & $-$49 08 08.9 & 0.542 & 213.1 & 17.72  & 15.92  & 13.75  & 2 & 11.54  & 11.00  & 10.61  ~& 23.6 & 10.8 & 11 & 17.6 & 2.8 & 12 & 45.1 & M5.5V &   3   & \\
SCR0113-7603     & 01 13 31.47 & $-$76 03 09.3 & 0.228 & 125.1 & 16.07  & 14.62  & 12.83  & 2 & 11.03  & 10.47  & 10.13  ~& 21.7 &  5.6 & 11 & 22.8 & 3.6 & 12 & 24.6 & ----- &   6   & \\      
SCR0135-6127     & 01 35 53.66 & $-$61 27 11.1 & 0.255 & 256.8 & 14.32  & 13.11  & 11.57  & 2 & 10.06  &  9.53  &  9.24  ~& 20.9 &  7.8 & 11 & 24.8 & 3.8 & 12 & 30.0 & M3.5V &   5   & \\
SCR0137-4148     & 01 37 23.49 & $-$41 48 56.2 & 0.238 & 054.3 & 15.33  & 13.99  & 12.30  & 2 & 10.68  & 10.07  &  9.78  ~& 21.8 &  6.3 & 11 & 24.1 & 3.8 & 12 & 27.2 & ----- &   6   & \\      
SCR0138-5353     & 01 38 20.51 & $-$53 53 26.1 & 0.297 & 071.0 & 14.36  & 13.20  & 11.70  & 2 & 10.28  &  9.69  &  9.42  ~& 24.3 &  7.4 & 11 & 29.7 & 4.6 & 12 & 41.8 & M3.5V &   5   & \\
SCR0150-3741     & 01 50 13.25 & $-$37 41 52.0 & 0.230 & 112.3 & 14.47  & 13.23  & 11.64  & 2 & 10.11  &  9.54  &  9.24  ~& 21.5 &  6.0 & 11 & 22.8 & 3.5 & 12 & 24.9 & ----- &   6   & \\      
SCR0211-6108     & 02 11 35.42 & $-$61 08 53.8 & 0.234 & 060.8 & 10.40  &  9.84  &  9.37  & 2 &  8.67  &  8.15  &  8.07  ~& 23.0 &  6.1 &  3 & 58.5 & 9.0 &  6 & 64.9 & K2.0V &   5   & \\
SCR0214-0733     & 02 14 44.69 & $-$07 33 24.7 & 0.399 & 128.9 &  9.89  &  9.35  &  8.88  & 2 &  8.20  &  7.80  &  7.60  ~& 21.2 &  8.3 &  7 & 47.3 & 7.3 &  6 & 89.5 & K1.0V &   6   & \tablenotemark {a}\\
SCR0232-8458     & 02 32 50.12 & $-$84 58 09.5 & 0.220 & 141.9 & 11.47  & 10.65  &  9.90  & 2 &  9.00  &  8.34  &  8.18  ~& 25.0 &  8.9 &  6 & 43.3 & 7.2 &  8 & 45.2 & K8.0V &   5   & \\  
SCR0246-7024     & 02 46 02.25 & $-$70 24 06.3 & 0.259 & 113.2 & 14.86  & 13.44  & 11.61  & 4 &  9.84  &  9.33  &  9.02  ~& 20.0 &  8.1 & 11 & 14.2 & 2.3 & 12 & 17.5 & M4.5V &   5   & \\      
SCR0325-0308     & 03 25 03.09 & $-$03 08 20.4 & 0.197 & 082.7 & 13.88  & 12.79  & 11.41  & 2 & 10.06  &  9.45  &  9.20  ~& 24.8 &  7.4 & 11 & 31.5 & 4.9 & 12 & 29.4 & ----- &   6   & \\      
SCR0420-7005     & 04 20 12.55 & $-$70 05 58.7 & 0.670 & 021.2 & 17.09  & 15.35  & 13.25  & 3 & 11.19  & 10.59  & 10.25  ~& 22.5 &  9.2 & 11 & 16.3 & 2.7 & 12 & 51.8 & M5.5V &  2,3  & \\
SCR0506-4712     & 05 06 07.29 & $-$47 12 51.6 & 0.207 & 319.4 & 12.63  & 11.65  & 10.46  & 2 &  9.32  &  8.66  &  8.43  ~& 21.1 &  5.7 & 11 & 30.2 & 4.9 & 12 & 29.7 & ----- &   6   & \\      
SCR0509-4325     & 05 09 43.85 & $-$43 25 17.4 & 0.225 & 324.9 & 14.09  & 12.83  & 11.21  & 2 &  9.61  &  9.00  &  8.74  ~& 18.0 &  5.2 & 11 & 16.5 & 2.6 & 12 & 17.6 & ----- &   6   & \\      
SCR0513-7653     & 05 13 05.97 & $-$76 53 21.9 & 0.274 & 301.2 & 12.99  & 11.89  & 10.52  & 2 &  9.24  &  8.63  &  8.36  ~& 16.2 &  8.6 & 11 & 22.6 & 3.6 & 12 & 29.4 & ----- &   6   & \\      
SCR0526-4851     & 05 26 40.80 & $-$48 51 47.3 & 0.279 & 158.7 & 12.88  & 11.79  & 10.43  & 2 &  9.10  &  8.46  &  8.24  ~& 20.1 &  5.3 & 11 & 20.7 & 3.3 & 12 & 27.3 & ----- &   6   & \\      
SCR0527-7231     & 05 27 06.99 & $-$72 31 20.0 & 0.368 & 018.3 & 14.71  & 13.49  & 11.86  & 3 & 10.34  &  9.76  &  9.47  ~& 22.7 &  6.5 & 11 & 25.1 & 3.9 & 12 & 43.8 & M3.5V &   5   & \\    
SCR0607-6115     & 06 07 58.09 & $-$61 15 10.5 & 0.217 & 008.5 & 14.57  & 13.26  & 11.57  & 2 &  9.98  &  9.38  &  9.10  ~& 19.8 &  5.9 & 11 & 18.4 & 2.9 & 12 & 18.9 & ----- &   6   & \\      
SCR0630-7643AB   & 06 30 46.61 & $-$76 43 08.9 & 0.483 & 356.8 & 14.82J & 13.08J & 11.00J & 4 &  8.89J &  8.28J &  7.92J ~&  6.9 &  2.0 & 11 &  5.5 & 0.9 & 12 & 12.5 & M5.0V &  2,3  & \tablenotemark {b} \tablenotemark {c}\\
SCR0631-8811     & 06 31 31.04 & $-$88 11 36.6 & 0.516 & 349.9 & 15.65  & 14.05  & 12.04  & 3 & 10.04  &  9.46  &  9.07  ~& 12.8 &  5.1 & 11 & 10.4 & 1.6 & 12 & 25.5 & M5.0V &   3   & \\               
SCR0635-6722     & 06 35 48.81 & $-$67 22 58.5 & 0.383 & 340.0 & 11.54  & 10.62  &  9.64  & 2 &  8.54  &  7.96  &  7.69  ~& 22.7 &  6.3 & 11 & 26.1 & 4.2 & 11 & 47.3 & M0.5V &   5   & \\        
SCR0640-0552     & 06 40 13.99 & $-$05 52 23.3 & 0.592 & 170.5 & 10.22  &  9.22  &  8.03  & 3 &  6.84  &  6.21  &  5.96  ~&  8.5 &  2.3 & 11 &  9.3 & 1.5 & 12 & 26.2 & M1.5V &   4   & \tablenotemark {d}\\   
SCR0642-6707     & 06 42 27.17 & $-$67 07 19.8 & 0.811 & 120.4 & 16.01  & 14.43  & 12.42  & 3 & 10.62  & 10.15  &  9.81  ~& 24.1 & 12.0 & 11 & 17.6 & 3.9 & 12 & 67.8 & M5.0V &   3   & \\        
SCR0643-7003     & 06 43 29.79 & $-$70 03 20.8 & 0.193 & 005.2 & 12.92  & 11.76  & 10.29  & 2 &  8.85  &  8.28  &  7.97  ~& 16.2 &  4.4 & 11 & 15.3 & 2.4 & 12 & 14.0 & ----- &   6   & \\      
SCR0644-4223AB   & 06 44 32.09 & $-$42 23 45.2 & 0.197 & 163.8 & 14.44J & 13.19J & 11.55J & 2 &  9.93J &  9.27J &  8.98J ~& 18.8 &  5.2 & 11 & 17.5 & 2.9 & 12 & 16.3 & ----- &   6   & \tablenotemark {e}\\      
SCR0702-6102     & 07 02 50.34 & $-$61 02 47.6 & 0.786 & 041.4 & 16.62  & 14.75  & 12.49  & 3 & 10.36  &  9.85  &  9.52  ~& 15.9 &  7.6 & 11 & 10.8 & 2.0 & 12 & 40.2 & M6.0V &  2,3  & \\
SCR0713-0511     & 07 13 11.23 & $-$05 11 48.6 & 0.304 & 183.6 & 11.13  & 10.08  &  8.86  & 3 &  7.65  &  7.08  &  6.82  ~& 13.1 &  4.7 & 10 & 13.5 & 2.1 & 12 & 19.4 & M1.5V &   6   & \\
SCR0717-0500     & 07 17 17.09 & $-$05 01 03.6 & 0.580 & 133.6 & 13.29  & 12.02  & 10.39  & 3 &  8.87  &  8.35  &  8.05  ~& 15.9 &  5.8 &  8 & 13.2 & 2.2 & 12 & 36.2 & M4.0V &   4   & \\
SCR0723-8015     & 07 23 59.65 & $-$80 15 17.8 & 0.828 & 330.4 & 17.41  & 15.61  & 13.41  & 2 & 11.30  & 10.82  & 10.44  ~& 19.3 &  5.6 & 11 & 17.1 & 3.1 & 12 & 67.2 & M5.5V &  2,3  & \\
SCR0736-3024     & 07 36 56.68 & $-$30 24 16.4 & 0.424 & 145.7 & 13.64  & 12.41  & 10.83  & 2 &  9.36  &  8.79  &  8.50  ~& 20.2 &  9.0 & 11 & 17.3 & 2.7 & 12 & 34.7 & M3.5V &   4   & \\
SCR0740-4257     & 07 40 11.80 & $-$42 57 40.3 & 0.714 & 318.1 & 13.81  & 12.36  & 10.50  & 3 &  8.68  &  8.09  &  7.77  ~& 10.0 &  2.9 & 11 &  7.2 & 1.1 & 12 & 24.5 & M4.5V &   4   & \tablenotemark {f}\\
SCR0754-3809     & 07 54 54.84 & $-$38 09 37.8 & 0.401 & 351.4 & 15.46  & 13.91  & 11.98  & 3 & 10.01  &  9.42  &  9.08  ~& 12.0 &  3.5 & 11 & 11.3 & 1.7 & 12 & 21.4 & M4.5V &   4   & \\
SCR0805-5912     & 08 05 46.17 & $-$59 12 50.4 & 0.637 & 155.0 & 14.69  & 13.38  & 11.68  & 3 & 10.07  &  9.52  &  9.22  ~& 20.4 &  6.0 & 11 & 19.4 & 3.1 & 12 & 58.7 & M4.0V &   3   & \\
SCR0838-5855     & 08 38 02.24 & $-$58 55 58.7 & 0.320 & 188.9 & 17.15  & 15.08  & 12.78  & 2 & 10.31  &  9.71  &  9.27  ~&  8.5 &  2.5 & 11 &  8.0 & 1.3 & 12 & 12.1 & M6.0V &   5   & \\
SCR0914-4134     & 09 14 17.42 & $-$41 34 37.9 & 0.749 & 312.5 & 15.01  & 13.57  & 11.72  & 3 &  9.98  &  9.42  &  9.12  ~& 18.2 &  8.0 & 11 & 14.6 & 2.4 & 12 & 51.9 & M4.5V &   4   & \\
SCR1107-3420B    & 11 07 50.25 & $-$34 21 00.6 & 0.287 & 167.0 & 15.04  & 13.68  & 11.96  & 3 & 10.26  &  9.70  &  9.41  ~& 19.2 &  5.5 & 11 & 19.1 & 3.0 & 12 & 26.0 & M4.0V &   6   & \tablenotemark {g}\tablenotemark {h}\\
SCR1110-3608     & 11 10 29.01 & $-$36 08 24.5 & 0.527 & 268.6 & 15.76  & 14.38  & 12.64  & 2 & 10.93  & 10.34  & 10.00  ~& 22.3 &  7.1 & 12 & 23.8 & 3.7 & 12 & 59.5 & ----- &   4   & \tablenotemark {i}\\
SCR1125-3834     & 11 25 37.27 & $-$38 34 43.0 & 0.586 & 252.1 & 14.57  & 13.29  & 11.67  & 2 & 10.09  &  9.51  &  9.19  ~& 18.1 &  5.4 & 11 & 20.5 & 3.2 & 12 & 56.8 & ----- &   4   & \tablenotemark {j}\\
SCR1138-7721     & 11 38 16.76 & $-$77 21 48.5 & 2.141 & 286.7 & 14.78  & 13.20  & 11.24  & 4 &  9.40  &  8.89  &  8.52  ~&  8.8 &  2.7 & 11 &  9.4 & 1.7 & 12 & 95.8 & M5.0V & 1,2,3 & \\
SCR1147-5504     & 11 47 52.49 & $-$55 04 11.9 & 0.192 & 011.3 & 13.72  & 12.54  & 11.06  & 2 &  9.67  &  9.08  &  8.81  ~& 24.2 &  8.1 & 11 & 23.1 & 3.6 & 12 & 21.0 & ------&   5   & \\
SCR1157-0149     & 11 57 45.55 & $-$01 49 02.6 & 0.451 & 116.4 & 15.99  & 14.54  & 12.68  & 2 & 10.91  & 10.35  & 10.02  ~& 22.2 &  6.5 & 11 & 21.3 & 3.4 & 12 & 45.6 & M4.5V &   4   & \\
SCR1204-4037     & 12 04 15.52 & $-$40 37 52.6 & 0.695 & 150.0 & 13.47  & 12.34  & 10.92  & 2 &  9.57  &  9.02  &  8.75  ~& 21.2 &  5.7 & 11 & 25.2 & 3.9 & 12 & 83.0 & ----- &   4   & \tablenotemark {k}\\
SCR1206-3500     & 12 06 58.52 & $-$35 00 52.0 & 0.422 & 229.3 & 14.67  & 13.35  & 11.66  & 2 & 10.01  &  9.40  &  9.13  ~& 21.0 &  5.9 & 11 & 17.7 & 2.7 & 12 & 35.4 & ----- &   4   & \\
SCR1214-4603     & 12 14 39.98 & $-$46 03 14.3 & 0.750 & 250.8 & 15.66  & 14.14  & 12.23  & 3 & 10.32  &  9.75  &  9.44  ~& 18.0 &  6.9 & 11 & 14.2 & 2.2 & 12 & 50.5 & ----- &   4   & \\ 
SCR1217-7810     & 12 17 26.93 & $-$78 10 45.9 & 0.212 & 056.6 & 16.43  & 14.92  & 13.05  & 2 & 11.20  & 10.64  & 10.36  ~& 24.5 &  8.6 & 11 & 23.3 & 3.8 & 12 & 23.4 & ----- &   5   & \\
SCR1220-8302     & 12 20 03.71 & $-$83 02 29.2 & 0.243 & 244.2 & 15.71  & 14.35  & 12.62  & 2 & 10.97  & 10.39  & 10.07  ~& 25.0 &  8.8 & 11 & 26.3 & 4.1 & 12 & 30.3 & ----- &   5   & \\ 
SCR1224-5339     & 12 24 24.44 & $-$53 39 08.8 & 0.189 & 251.9 & 15.31  & 13.95  & 12.19  & 2 & 10.51  &  9.93  &  9.65  ~& 18.2 &  5.5 & 11 & 21.0 & 3.3 & 12 & 18.8 & ----- &   5   & \\ 
SCR1230-3411     & 12 30 01.75 & $-$34 11 24.1 & 0.527 & 234.9 & 14.16  & 12.81  & 11.07  & 3 &  9.34  &  8.77  &  8.44  ~& 12.6 &  3.7 & 11 & 11.7 & 1.8 & 12 & 29.3 & ----- &   4   & \tablenotemark {l}\\
SCR1240-8116     & 12 40 56.02 & $-$81 16 31.0 & 0.492 & 279.8 & 14.11  & 12.89  & 11.28  & 3 &  9.73  &  9.16  &  8.89  ~& 19.2 &  6.1 & 11 & 19.2 & 2.9 & 12 & 44.7 & M4.0V &   3   & \\
SCR1245-5506     & 12 45 52.53 & $-$55 06 50.2 & 0.412 & 107.0 & 13.66  & 12.32  & 10.61  & 3 &  8.99  &  8.43  &  8.12  ~& 11.5 &  3.4 & 11 & 11.3 & 1.8 & 12 & 22.1 & M4.0V &   3   & \\
SCR1247-0525     & 12 47 14.73 & $-$05 25 13.2 & 0.722 & 319.8 & 14.77  & 13.44  & 11.72  & 2 & 10.13  &  9.62  &  9.29  ~& 24.2 &  9.6 & 11 & 20.3 & 3.5 & 12 & 69.4 & M4.0V &   4   & \\
SCR1347-7610     & 13 47 56.80 & $-$76 10 20.0 & 0.194 & 089.7 & 11.49  & 10.57  &  9.66  & 3 &  8.62  &  8.01  &  7.77  ~& 22.6 &  6.2 & 11 & 29.2 & 4.7 &  9 & 26.8 & ----- &   5   & \\ 
SCR1420-7516     & 14 20 36.84 & $-$75 16 05.9 & 0.195 & 243.7 & 13.78  & 12.55  & 10.95  & 2 &  9.44  &  8.91  &  8.63  ~& 21.4 &  6.7 & 11 & 17.9 & 2.9 & 12 & 16.6 & ----- &   5   & \\ 
SCR1441-7338     & 14 41 14.42 & $-$73 38 41.4 & 0.207 & 029.0 & 16.96  & 15.31  & 13.25  & 2 & 11.20  & 10.61  & 10.27  ~& 19.0 &  5.5 & 11 & 17.2 & 2.7 & 12 & 16.8 & ----- &   5   & \\ 
SCR1444-3426     & 14 44 06.56 & $-$34 26 47.2 & 0.451 & 187.7 & 14.17  & 12.89  & 11.26  & 2 &  9.74  &  9.18  &  8.88  ~& 24.0 &  7.5 & 11 & 18.8 & 3.0 & 12 & 40.1 & ----- &   4   & \tablenotemark {m}\\ 
SCR1448-5735     & 14 48 39.82 & $-$57 35 17.7 & 0.202 & 188.8 & 11.15  & 10.47  &  9.92  & 2 &  9.15  &  8.56  &  8.43  ~& 18.3 & 17.5 &  6 & 62.4 & 9.7 &  6 & 59.7 & ----- &   5   & \\
SCR1450-3742     & 14 50 02.85 & $-$37 42 09.8 & 0.449 & 212.2 & 14.01  & 12.84  & 11.32  & 2 &  9.95  &  9.37  &  9.07  ~& 21.2 &  6.0 & 11 & 25.9 & 4.2 & 12 & 55.1 & ----- &   4   & \tablenotemark {n}\\ 
SCR1456-7239     & 14 56 02.29 & $-$72 39 41.4 & 0.207 & 225.0 & 15.36  & 13.99  & 12.28  & 2 & 10.62  & 10.06  &  9.74  ~& 24.9 &  7.0 & 11 & 22.8 & 3.6 & 12 & 22.4 & ----- &   5   & \\ 
SCR1511-3403     & 15 11 38.63 & $-$34 03 16.6 & 0.561 & 202.9 & 14.45  & 13.22  & 11.60  & 2 & 10.05  &  9.42  &  9.13  ~& 16.1 &  7.4 & 11 & 20.5 & 3.2 & 12 & 54.4 & ----- &   4   & \tablenotemark {o}\\ 
SCR1532-3622     & 15 32 13.90 & $-$36 22 30.9 & 0.438 & 235.4 & 14.03  & 12.89  & 11.45  & 2 & 10.10  &  9.54  &  9.28  ~& 23.0 & 10.3 & 11 & 31.5 & 5.0 & 12 & 65.4 & ----- &   4   & \tablenotemark {p}\\ 
SCR1601-3421     & 16 01 55.68 & $-$34 21 56.8 & 0.683 & 118.3 & 16.18  & 14.71  & 12.86  & 2 & 10.96  & 10.33  &  9.98  ~& 20.2 & 11.9 & 11 & 18.7 & 3.0 & 12 & 60.4 & ----- &   4   & \tablenotemark {q}\\
SCR1630-3633AB   & 16 30 27.23 & $-$36 33 56.1 & 0.413 & 249.2 & 14.95J & 13.47J & 11.59J & 2 & 10.04J &  9.50J &  9.03J ~& 14.8 &  5.9 & 11 & 15.4 & 3.8 & 12 & 30.1 & ----- &   4   & \tablenotemark {r}\\
SCR1637-4703     & 16 37 56.62 & $-$47 03 44.4 & 0.503 & 215.4 & 14.77  & 13.57  & 12.04  & 3 & 10.60  & 10.04  &  9.70  ~& 20.8 & 13.7 &  9 & 32.0 & 5.1 & 12 & 76.2 & ----- &   4   & \\
SCR1726-8433     & 17 26 22.90 & $-$84 33 08.2 & 0.518 & 134.8 & 14.25  & 13.00  & 11.41  & 2 &  9.87  &  9.33  &  9.02  ~& 20.1 &  5.5 & 11 & 20.6 & 3.2 & 12 & 50.6 & M4.0V &   3   & \\
SCR1738-5942     & 17 38 41.02 & $-$59 42 24.4 & 0.280 & 148.2 & 14.95  & 13.65  & 11.96  & 2 & 10.38  &  9.83  &  9.58  ~& 20.8 &  6.0 & 11 & 24.0 & 4.0 & 12 & 31.9 & ----- &   5   & \\
SCR1746-8211     & 17 46 21.54 & $-$82 11 56.6 & 0.228 & 184.9 & 12.44  & 11.33  &  9.91  & 2 &  8.55  &  7.99  &  7.71  ~& 14.6 &  4.0 & 11 & 15.5 & 2.4 & 12 & 16.7 & ----- &   5   & \tablenotemark {s}\\
SCR1820-6225     & 18 20 49.35 & $-$62 25 52.7 & 0.190 & 164.8 & 12.04  & 11.12  & 10.17  & 2 &  9.14  &  8.49  &  8.30  ~& 22.5 &  9.5 &  8 & 36.6 & 5.8 & 10 & 33.0 & ----- &   5   & \\
SCR1826-6542     & 18 26 46.83 & $-$65 42 39.9 & 0.311 & 178.9 & 17.35  & 15.28  & 12.96  & 3 & 10.57  &  9.96  &  9.55  ~&  9.2 &  2.5 & 11 &  9.3 & 1.5 & 12 & 13.7 & ----- &   5   & \\
SCR1841-4347     & 18 41 09.81 & $-$43 47 32.8 & 0.790 & 264.2 & 16.46  & 14.72  & 12.59  & 3 & 10.48  &  9.94  &  9.60  ~& 14.6 &  4.3 & 11 & 11.9 & 2.0 & 12 & 44.5 & ----- &   4   & \\
SCR1845-6357AB   & 18 45 05.26 & $-$63 57 47.8 & 2.558 & 074.7 & 17.40J & 14.99J & 12.46J & 5 &  9.54J &  8.97J &  8.51J ~&  3.5 &  1.2 &  6 &  4.6 & 0.8 & 10 & 56.3 & M8.5V & 1,2,3 & \tablenotemark {t}\tablenotemark {u}\\
SCR1847-1922     & 18 47 16.69 & $-$19 22 20.8 & 0.626 & 230.7 & 14.35  & 13.11  & 11.48  & 2 &  9.91  &  9.38  &  9.09  ~& 23.0 &  6.6 & 11 & 20.6 & 3.2 & 12 & 61.1 & ----- &   4   & \\
SCR1853-7537     & 18 53 26.61 & $-$75 37 39.8 & 0.304 & 168.7 & 11.18  & 10.29  &  9.41  & 2 &  8.34  &  7.73  &  7.50  ~& 20.1 &  6.1 & 11 & 25.9 & 4.4 & 10 & 37.3 & ----- &   5   & \tablenotemark {v}\\
SCR1855-6914     & 18 55 47.89 & $-$69 14 15.1 & 0.832 & 145.3 & 16.61  & 14.79  & 12.66  & 3 & 10.47  &  9.88  &  9.51  ~& 12.5 &  4.5 & 11 & 10.7 & 1.7 & 12 & 42.1 & M5.5V &   3   & \\
SCR1856-4704AB   & 18 56 38.40 & $-$47 04 58.3 & 0.252 & 131.3 & 14.85J & 13.55J & 11.87J & 2 & 10.29J &  9.75J &  9.45J ~& 21.4 &  6.2 & 11 & 22.6 & 3.6 & 12 & 27.0 & ----- &   5   & \tablenotemark {w}\\
SCR1926-8216AB   & 19 26 48.64 & $-$82 16 47.6 & 0.195 & 172.5 & 11.29J & 10.55J &  9.89J & 3 &  9.04J &  8.43J &  8.31J ~& 17.3 &  4.7 &  3 & 53.4 & 8.5 &  6 & 49.3 & ----- &   5   & \tablenotemark {x}\\
SCR1931-0306     & 19 31 04.61 & $-$03 06 18.0 & 0.578 & 031.0 & 16.81  & 15.11  & 13.11  & 3 & 11.15  & 10.56  & 10.23  ~& 18.0 &  4.8 &  6 & 18.0 & 3.0 & 12 & 49.2 & M5.0V &   4   & \\
SCR1932-0652     & 19 32 46.33 & $-$06 52 18.1 & 0.318 & 193.3 & 13.97  & 12.83  & 11.34  & 2 &  9.94  &  9.36  &  9.10  ~& 23.7 &  9.1 & 11 & 26.7 & 4.1 & 12 & 40.2 & ----- &   6   & \\
SCR1932-5005     & 19 32 48.64 & $-$50 05 38.9 & 0.257 & 157.5 & 15.47  & 14.14  & 12.42  & 2 & 10.75  & 10.11  &  9.85  ~& 24.4 &  7.8 & 11 & 23.5 & 3.7 & 12 & 28.6 & ----- &   5   & \\
SCR1959-3631     & 19 59 21.03 & $-$36 31 03.7 & 0.436 & 158.1 & 11.07  & 10.19  &  9.31  & 2 &  8.24  &  7.62  &  7.41  ~& 19.8 &  5.4 &  8 & 25.0 & 4.3 & 10 & 51.6 & ----- &   4   & \tablenotemark {y}\\
SCR1959-6236     & 19 59 33.55 & $-$62 36 13.4 & 0.189 & 288.7 & 16.27  & 14.82  & 12.96  & 2 & 11.07  & 10.49  & 10.23  ~& 24.4 &  7.2 & 11 & 21.9 & 3.4 & 12 & 19.6 & ----- &   5   & \\
SCR2009-0113     & 20 09 18.24 & $-$01 13 38.2 & 0.381 & 186.2 & 14.47  & 12.98  & 11.16  & 3 &  9.40  &  8.83  &  8.51  ~& 14.0 &  4.4 & 11 & 10.8 & 1.8 & 12 & 19.5 & ----- &   6   & \\
SCR2016-7531     & 20 16 11.25 & $-$75 31 04.5 & 0.253 & 081.3 & 15.84  & 14.28  & 12.35  & 3 & 10.47  &  9.86  &  9.51  ~& 16.2 &  4.8 & 11 & 14.3 & 2.3 & 12 & 17.1 & ----- &   5   & \tablenotemark {z}\\
SCR2018-3635     & 20 18 06.52 & $-$36 35 27.7 & 0.237 & 125.0 & 14.63  & 13.38  & 11.75  & 2 & 10.21  &  9.67  &  9.44  ~& 20.7 &  6.9 & 11 & 24.9 & 4.1 & 12 & 28.0 & ----- &   6   & \\      
SCR2025-2259     & 20 25 18.93 & $-$22 59 06.0 & 0.191 & 219.5 & 14.22  & 13.03  & 11.48  & 2 & 10.00  &  9.48  &  9.16  ~& 23.5 &  7.0 & 11 & 24.6 & 3.8 & 12 & 22.2 & ----- &   6   & \\      
SCR2033-4903     & 20 33 01.87 & $-$49 03 10.6 & 0.261 & 150.4 & 15.34  & 13.85  & 11.98  & 2 & 10.11  &  9.52  &  9.19  ~& 16.5 &  6.4 & 11 & 13.2 & 2.1 & 12 & 16.4 & ----- &   6   & \\      
SCR2040-5501     & 20 40 12.38 & $-$55 01 25.6 & 0.514 & 125.4 & 15.22  & 13.91  & 12.22  & 2 & 10.56  & 10.02  &  9.69  ~& 22.9 &  6.9 & 11 & 23.3 & 3.6 & 12 & 56.8 & M4.0V &   3   & \\
SCR2042-5737AB   & 20 42 46.44 & $-$57 37 15.3 & 0.264 & 142.6 & 14.12J & 12.89J & 11.33J & 3 &  9.97J &  9.53J &  9.03J ~& 22.7 &  8.5 & 11 & 25.3 & 6.0 & 12 & 31.6 & ----- &   5   & \tablenotemark {A}\\
SCR2104-5248     & 21 04 53.85 & $-$52 48 34.3 & 0.326 & 182.7 & 13.66  & 12.52  & 11.08  & 2 &  9.72  &  9.10  &  8.83  ~& 22.1 &  6.4 & 11 & 24.5 & 3.8 & 12 & 37.9 & ----- &   6   & \\      
SCR2105-5503     & 21 05 13.78 & $-$55 03 56.3 & 0.334 & 171.0 & 13.97  & 12.69  & 11.05  & 2 &  9.59  &  8.92  &  8.64  ~& 17.1 &  4.8 & 11 & 16.7 & 2.8 & 12 & 26.4 & ----- &   6   & \\      
SCR2122-4314     & 21 22 16.92 & $-$43 14 05.0 & 0.262 & 184.7 & 13.39  & 12.17  & 10.61  & 3 &  9.13  &  8.53  &  8.21  ~& 17.0 &  4.7 & 11 & 14.9 & 2.3 & 12 & 18.5 & ----- &   6   & \\      
SCR2130-7710     & 21 30 07.00 & $-$77 10 37.5 & 0.589 & 118.0 & 17.01  & 15.33  & 13.31  & 2 & 11.29  & 10.67  & 10.37  ~& 20.6 &  7.3 & 11 & 18.4 & 2.9 & 12 & 51.3 & M4.5V &   3   & \tablenotemark {B}\\
SCR2135-5325     & 21 35 39.63 & $-$53 25 31.5 & 0.191 & 332.4 & 15.88  & 14.39  & 12.54  & 2 & 10.80  & 10.25  &  9.92  ~& 20.8 &  6.7 & 11 & 20.7 & 3.6 & 12 & 18.8 & ----- &   6   & \\      
SCR2142-7405     & 21 42 58.54 & $-$74 05 55.5 & 0.210 & 162.1 & 13.06  & 12.02  & 10.74  & 2 &  9.46  &  8.78  &  8.58  ~& 21.9 &  6.3 & 11 & 26.7 & 4.5 & 12 & 26.5 & ----- &   6   & \\      
SCR2230-5244     & 22 30 27.95 & $-$52 44 29.1 & 0.369 & 125.7 & 17.37  & 15.91  & 13.88  & 2 & 11.85  & 11.24  & 10.91  ~& 24.7 &  7.1 &  6 & 24.8 & 4.2 & 12 & 43.4 & ----- &   5   & \\
SCR2241-6119A    & 22 41 44.36 & $-$61 19 31.2 & 0.184 & 124.0 & 14.41  & 13.21  & 11.66  & 2 & 10.21  &  9.61  &  9.35  ~& 23.2 &  7.3 & 11 & 26.6 & 4.1 & 12 & 23.2 & ----- &   5   & \\
SCR2241-6119B    & 22 41 43.67 & $-$61 19 39.4 & 0.287 & 108.0 & 19.15  & 17.23  & 14.96  & 2 & 12.64  & 12.05  & 11.68  ~& 27.8 & 16.5 & 11 & 26.2 & 4.1 & 12 & 35.6 & ----- &   5   & \tablenotemark {C}\\
SCR2252-2220     & 22 52 25.82 & $-$22 20 06.8 & 0.299 & 187.6 & 13.71  & 12.56  & 11.10  & 2 &  9.70  &  9.11  &  8.86  ~& 21.3 &  5.7 & 11 & 24.1 & 3.7 & 12 & 34.2 & ----- &   6   & \\      
SCR2301-5530     & 23 01 32.51 & $-$55 30 17.6 & 0.338 & 052.8 & 12.65  & 11.57  & 10.25  & 2 &  8.98  &  8.36  &  8.13  ~& 15.1 &  4.5 & 11 & 21.4 & 3.3 & 12 & 34.3 & ----- &   6   & \\      
SCR2307-8452     & 23 07 19.67 & $-$84 52 03.9 & 0.613 & 097.2 & 15.13  & 13.76  & 12.00  & 2 & 10.36  &  9.81  &  9.47  ~& 20.6 &  5.9 & 11 & 19.9 & 3.2 & 12 & 57.8 & M4.0V &   3   & \\
SCR2325-6740     & 23 25 25.13 & $-$67 40 07.9 & 0.284 & 123.2 & 14.38  & 13.11  & 11.48  & 2 &  9.91  &  9.33  &  9.05  ~& 20.0 &  5.7 & 11 & 19.5 & 3.0 & 12 & 26.2 & ----- &   6   & \tablenotemark {D}\\    
SCR2335-6433A    & 23 35 18.43 & $-$64 33 42.4 & 0.196 & 103.1 & 11.20  & 10.42  &  9.64  & 2 &  8.64  &  8.02  &  7.86  ~& 24.5 &  6.9 &  8 & 35.9 & 6.4 &  9 & 33.3 & ----- &   5   & \\
SCR2335-6433B    & 23 35 19.95 & $-$64 33 22.3 & 0.196 & 099.1 & 15.65  & 14.49  & 13.00  & 2 & 11.60  & 11.02  & 10.76  ~& 57.6 & 15.4 & 11 & 56.7 & 8.8 & 12 & 52.7 & ----- &   5   & \tablenotemark {E}\\

\enddata

\tablecomments{coordinates are epoch and equinox 2000.0 using
SuperCOSMOS positions transformed to epoch 2000.0 using the proper
motions and position angles listed here; if AB, photometry or spectral
type is for the system, i.e. joint.}
  
\tablecomments{(1): {Hambly et al.~2004}; (2): {Henry et al.~2004};
(3): {Subasavage et al.~2005a}; (4): {Subasavage et al.~2005b}; (5):
{Finch et al.~2007}; (6): {this paper}.}
\tablenotetext {a} {~BD $-$08 409}
\tablenotetext {b} {~AB separation 1\arcsec}
\tablenotetext {c} {~SIPS J0630-7643}
\tablenotetext {d} {~BD $-$05 1737}
\tablenotetext {e} {~angular separation of 1.6\arcsec~at a position angle of 266.4$^\circ$}
\tablenotetext {f} {~LSR J07401-4257}
\tablenotetext {g} {~SCR1107-3420A is a white dwarf.}
\tablenotetext {h} {~angular separation of 30.6\arcsec~at a position angle of 107.1$^\circ$}
\tablenotetext {i} {~LSR J11104-3608}
\tablenotetext {j} {~LSR J11256-3834}
\tablenotetext {k} {~LSR J12042-4037}
\tablenotetext {l} {~LSR J12300-3411}
\tablenotetext {m} {~LSR J14441-3426}
\tablenotetext {n} {~LSR J14500-3742}
\tablenotetext {o} {~LSR J15116-3403}
\tablenotetext {p} {~LSR J15322-3622}
\tablenotetext {q} {~LSR J16019-3421}
\tablenotetext {r} {~angular separation of 2\arcsec~at a position angle of 247.2$^\circ$}
\tablenotetext {s} {~possible companion to HD158866 at 76.5\arcsec~at a position angle of 290.8$^\circ$}
\tablenotetext {t} {~DENIS-P J184504.9-635747}
\tablenotetext {u} {~angular separation of 1.2\arcsec~at a position angle of 170.2$^\circ$}
\tablenotetext {v} {~USNO-B1.0 0143-00184491}
\tablenotetext {w} {~angular separation of 1.1\arcsec~at a position angle of 139.9$^\circ$}
\tablenotetext {x} {~angular separation of 1.4\arcsec~at a position angle of 238.8$^\circ$}
\tablenotetext {y} {~CD $-$36 13808}
\tablenotetext {z} {~SIPS J2016-7531}
\tablenotetext {A} {~angular separation of 2.2\arcsec~at a position angle of 338.4$^\circ$}
\tablenotetext {B} {~SIPS J2130-7710}
\tablenotetext {C} {~angular separation of 9.6\arcsec~at a position angle of 211.2$^\circ$}
\tablenotetext {D} {~LEHPM 5810}
\tablenotetext {E} {~angular separation of 22.4\arcsec~at a position angle of 25.9$^\circ$}

\end{deluxetable}
}


\voffset0pt{
\centering
\begin{deluxetable}{lrrrrc}
\tabletypesize{\scriptsize}
\tablecaption{Binary Data for SCR Sample}
\setlength{\tabcolsep}{0.03in}
\label{tab:bindata}
\tablewidth{0pt}

\tablehead{\colhead{Name}                &
           \colhead{sep}                 &
           \colhead{P.A.}                &
           \colhead{d$_{ccd}$}           &
           \colhead{$\sigma$$_{ccd}$}    &
           \colhead{notes}               \\

           \colhead{   }                 &
           \colhead{(\arcsec)}           &
           \colhead{(deg)}               &
           \colhead{(pc)}                &
           \colhead{(pc)}                &
           \colhead{}                    }

\startdata
SCR0630-7643A   &      &       &   8.8  & 0.9 &  a     \\
SCR0630-7643B   &  1.0 &  33.8 &  [5.5] & 0.9 &        \\
\hspace*{\fill}                                        \\
SCR0644-4223A   &      &       & [17.5] & 2.9 &        \\
SCR0644-4223B   &  1.6 & 266.4 & [17.5] & 2.9 &        \\
\hspace*{\fill}                                        \\
SCR1107-3420A   &      &       &  28.2  & 3.0 &  b     \\
SCR1107-3420B   & 30.6 & 107.1 &  19.1  & 3.0 &  b     \\
\hspace*{\fill}                                        \\
SCR1630-3633A   &      &       & [15.4] & 3.8 &        \\
SCR1630-3633B   &  2.0 & 247.2 & [15.4] & 3.8 &        \\
\hspace*{\fill}                                        \\
HD158866        &      &       &  30.6  & 0.9 &  c     \\
SCR1746-8211    & 76.5 & 290.8 &  15.5  & 2.4 &  c     \\
\hspace*{\fill}                                        \\
SCR1845-6357A   &      &       &   3.9  & 0.8 &  a     \\
SCR1845-6357B   &  1.2 & 170.2 &  [4.6] & 0.8 &  d     \\
\hspace*{\fill}                                        \\
SCR1856-4704A   &      &       & [22.6] & 3.6 &        \\
SCR1856-4704B   &  1.1 & 138.9 & [22.6] & 3.6 &        \\
\hspace*{\fill}                                        \\
SCR1926-8216A   &      &       & [53.4] & 8.5 &        \\
SCR1926-8216B   &  1.4 & 238.8 & [53.4] & 8.5 &        \\
\hspace*{\fill}                                        \\
SCR2042-5737A   &      &       & [25.3] & 6.0 &        \\
SCR2042-5737B   &  2.2 & 338.4 & [25.3] & 6.0 &        \\
\hspace*{\fill}                                        \\
SCR2241-6119A   &      &       &  26.6  & 4.1 &  e     \\
SCR2241-6119B   &  9.6 & 211.2 &  26.2  & 4.1 &  e     \\
\hspace*{\fill}                                        \\
SCR2335-6433A   &      &       &  35.9  & 6.4 &  e     \\
SCR2335-6433B   & 22.4 &  25.9 &  56.7  & 8.8 &  e     \\
\enddata

\tablecomments{Brackets around the distance estimate indicate that the
estimate is for the system}

\tablenotetext {a} {trigonometric parallax in Henry et al.~2006}  

\tablenotetext {b} {separation and position angle for the system and
distance estimate for the primary (derived from the SED) are from
Subasavage et al.~2007}

\tablenotetext {c} {separation and position angle for the system from
Finch et al.~2007; trigonometric distance for the primary is from ~van
Leeuwen 2007 (HIPPARCOS)}

\tablenotetext {d} {separation and position angle for the system from
Biller et al.~2006}

\tablenotetext {e} {separation and position angle for the system from
Finch et al.~2007}

\end{deluxetable}
}


\voffset0pt{
\centering
\begin{deluxetable}{lcc}
\tabletypesize{\scriptsize}
\tablecaption{Distance Statistics for SCR Sample of M Dwarf Systems}
\setlength{\tabcolsep}{0.03in}
\label{tab:distdata}
\tablewidth{0pt}

\tablehead{\colhead{Distance}                &
           \colhead{\# systems}              &
           \colhead{\# binaries}\tablenotemark{a}}

\startdata
d $\leq$ 5.0 pc            &  1  &  1 \\
\hspace*{\fill}                       \\
5.0 $<$ d $\leq$ 10.0 pc   &  6  &  1 \\
\hspace*{\fill}                       \\
10.0 $<$ d $\leq$ 15.0 pc  & 16  &  0 \\
\hspace*{\fill}                       \\
15.0 $<$ d $\leq$ 20.0 pc  & 23  &  2 \\
\hspace*{\fill}                       \\
20.0 $<$ d $\leq$ 25.0 pc  & 31  &  1 \\
\hspace*{\fill}                       \\
d $\geq$ 25.0 pc           & 23  &  4 \\
\hline
\hspace*{\fill}                       \\
TOTAL                      &100 &   9 \\
\enddata

\tablenotetext {a} {omitting SCR1107-3420B and SCR1746-8211, both of
which have primaries that are not M dwarfs}

\end{deluxetable}
}

\end{document}